\documentclass[reprint,twocolumn,superscriptaddress]{revtex4}

\pdfoutput=1

\usepackage[utf8x]{inputenc}
\usepackage[english]{babel}
\usepackage{amsmath,amssymb,bm,wasysym}
\usepackage{graphicx}

\newcommand{\gam}{\bm{\gamma}}
\newcommand{\Gam}{\bm{\Gamma}}
\newcommand{\A}{{A}}
\newcommand{\B}{{B}}
\renewcommand{\d}{\mbox{d}}
\newcommand{\e}{\mbox{e}}
\newcommand{\ptil}{\widetilde{p}}
\newcommand{\vtil}{\widetilde{V}}
\newcommand{\papp}{p_{app}}

\newcommand{\meanc}[1]{{\left\langle #1 \right\rangle}_c}

\usepackage{ulem}
\usepackage{color}

\newcommand{\am}[1]{\textcolor{black}{#1}}

\DeclareRobustCommand{\am}{\textcolor{black}}

\begin{document}

\title{Towards a scale-bridging description of ferrogels and magnetic elastomers}

\author{Giorgio Pessot}
\email{giorgpess@thphy.uni-duesseldorf.de}
\affiliation{Institut f\"ur Theoretische Physik II: Weiche Materie,
 Heinrich-Heine-Universit\"at D\"usseldorf, D-40225 D\"usseldorf, Germany}

\author{Rudolf Weeber}
\email{weeber@icp.uni-stuttgart.de}
\affiliation{Institute for Computational Physics,
Universit\"at Stuttgart, 70569 Stuttgart, Germany}

\author{Christian Holm}
\email{holm@icp.uni-stuttgart.de}
\affiliation{Institute for Computational Physics,
Universit\"at Stuttgart, 70569 Stuttgart, Germany}

\author{Hartmut L\"owen}
\email{hlowen@thphy.uni-duesseldorf.de}
\affiliation{Institut f\"ur Theoretische Physik II: Weiche Materie,
 Heinrich-Heine-Universit\"at D\"usseldorf, D-40225 D\"usseldorf, Germany}

\author{Andreas M.\ Menzel}
\email{menzel@thphy.uni-duesseldorf.de}
\affiliation{Institut f\"ur Theoretische Physik II: Weiche Materie,
 Heinrich-Heine-Universit\"at D\"usseldorf, D-40225 D\"usseldorf, Germany}

\date{\today}

\begin{abstract}
Ferrogels and magnetic elastomers differentiate themselves from other materials by their unique
 capability of reversibly changing shape and mechanical properties under the influence
 of an external magnetic field.
A crucial issue in the study of these outstanding materials is the interaction between
 the mesoscopic magnetic particles and the polymer matrix in which they are embedded.
Here we analyze interactions between two such particles connected by a polymer chain,
 a situation representative for particle-crosslinked magnetic gels.
To make a first step towards a scale-bridging description of the materials,
 effective \am{pair} potentials for mesoscopic configurational changes are specified using microscopic
  input obtained from simulations.
Furthermore, the impact of the presence of magnetic interactions on the probability distributions and
 thermodynamic quantities of the system is considered.
The resulting mesoscopic model \am{pair} potentials can be used to economically model the system
 on the particle length scales.
This first coarse-graining step is important to realize simplified but realistic
 scale-bridging models for these promising materials.
\end{abstract}


\maketitle

\section{Introduction}\label{intro}
Ferrogels and magnetic elastomers are fascinating materials, born by the union of polymeric networks
 and ferrofluids.
Their amazing properties derive from the unique combination of the elastic behavior typical for
 polymers and rubbers \cite{strobl1997physics} on the one hand, and magnetic effects characteristic
 of ferrofluids and magnetorheological fluids \cite{klapp2005dipolar,huke2004magnetic,
rosensweig1985ferrohydrodynamics,odenbach2003ferrofluids,odenbach2003magnetoviscous,
odenbach2004recent,ilg2005structure,ilg2006structure,holm2005structure} on the other.
One of the most interesting results is that their shape and mechanical properties can be externally
 controlled by applying a magnetic field \cite{menzel2014tuned,jarkova2003hydrodynamics,zrinyi1996deformation,
deng2006development,stepanov2007effect, filipcsei2007magnetic,guan2008magnetostrictive,
bose2009magnetorheological, gong2012full,evans2012highly,borin2013tuning}.
A form of tunability, distinguished by reversibility as well as non-invasiveness and based on a
  magneto-mechanical coupling is one of the most appealing properties of these materials.
This makes them excellent candidates for the use as soft actuators \cite{zimmermann2006modelling},
 magnetic field detectors \cite{szabo1998shape,ramanujan2006mechanical}, as well as
 tunable vibration and shock absorbers \cite{deng2006development,sun2008study}.
Moreover, studying their heat dissipation due to hysteretic remagnetization in an alternating external magnetic
 field might be helpful to understand better the processes during possible applications in hyperthermal cancer
 treatment \cite{babincova2001superparamagnetic,lao2004magnetic}.

Typically, these materials consist of cross-linked polymer networks in which magnetic particles of
 nano- or micrometer size are dispersed \cite{filipcsei2007magnetic}.
A central role in the coupling of magnetic and mechanical properties is played by the specific interactions
 between the embedded mesoscopic magnetic particles and the flexible polymer chains filling the space
 between them.
These couplings are responsible for a modified macroscopic elasticity \cite{landau1975elasticity,
jarkova2003hydrodynamics,bohlius2004macroscopic}, orientational memory effects
 \cite{annunziata2013hardening,messing2011cobalt}, and reversibility of the magnetically induced
 deformations \cite{stepanov2007effect}.

Many theoretical and computational studies have been performed on the topic, using different
 approaches to incorporate the particle-polymer interaction.
Some of them rely on a continuum-mechanical description of both the polymeric matrix
 and the magnetic component \cite{szabo1998shape,zubarev2012theory,brand2014macroscopic}.
  Others explicitly take into account the discrete embedded magnetic particles, but treat the
 polymer matrix as an elastic background continuum \cite{ivaneyko2012effects,wood2011modeling}.
Usually in these studies, an affine deformation of the whole sample is assumed.
The limitations of such an approach for the characterization of real materials have
 recently been pointed out \cite{pessot2014structural}.
In order to include irregular distributions of particles and non-affine
 sample deformations, other works employ, for instance, finite-element methods
 \cite{raikher2008shape,stolbov2011modelling,gong2012full,han2013field,spieler2013xfem}.

To optimize economical efficiency, a first step is the use of simplified dipole-spring models.
In this case, steric repulsion and other effects like orientational memory terms can be included
 \cite{annunziata2013hardening,dudek2007molecular,sanchez2013effects,
tarama2014tunable,pessot2014structural,cerda2013phase}.
A step beyond the often used harmonic spring potentials can be found in \cite{sanchez2013effects}
 where non-linear springs of finite extensibility are considered.

From all the studies mentioned above it becomes clear that microscopic approaches that explicitly resolve polymer chains are rare \cite{weeber2012deformation,weeber2015ferrogels}, and what is particularly missing is a link between such microscopic approaches and the mesoscopic models that only resolve the magnetic particles, not the single polymer chains.
In particular, a microscopic foundation of the phenomenological mesoscopic expressions for the model energies should be built up.

The present work is a first step to close this gap.
We consider an explicit microscopic description in a first simplified approach: a single polymer chain, discretized through multiple beads, each representing a coarse-grained small part of the polymer, connects two mesoscopic particles.
The ends of the polymer chain are rigidly anchored on the surfaces of the two mesoscopic particles,
 which are spherical, can be magnetized, and are free to rotate and change their distance.
From molecular dynamics simulations on the microscopic level, we collect the statistics of the micro-states
 corresponding to different configurations in the mesoscopic model.
Based on these statistics we derive effective mesoscopic \am{pair} potentials, inspired by previous achievements for other polymeric systems \cite{harmandaris2007comparison,mulder2008equilibration}.
The subsequent step in scale-bridging, connecting the mesoscopic picture to the macroscopic description, has been recently addressed \cite{menzel2014bridging} for a special class of magnetic polymeric materials.

In the following, we first define and describe our model in section \ref{sys}.
Then, in section \ref{micro}, we mention the details of the microscopic simulation.
In section \ref{stat}, we further characterize the probability distribution of the mesoscopic variables, connecting it to a wrapping effect in section \ref{wrap}.
After that, in section \ref{margpot}, we determine the values of mesoscopic model parameters based on
 the results of our microscopic simulations.
In section \ref{effpot}, we derive an approximated expression for the mesoscopic effective \am{pair} potential characterizing the particle configurations.
In this way, we build the bridge from the explicit microscopic characterization to the mesoscopic particle-resolved models by averaging over the microscopic details. 
Last, in section \ref{magneff}, we consider the effect of adding magnetic moments to the particles and, in section \ref{thermod}, the effect of increasing magnetic interaction on the thermodynamic properties, before we draw our final conclusions in section \ref{conclusions}.
\am{Appendix~\ref{app2} addresses the trends in the dependences of the mesoscopic model parameters on varying microscopic system parameters, while appendix~\ref{app1} briefly comments on the separability of magnetic interactions in the mesoscopic picture and microscopic chain configurations.}

\section{The System}\label{sys}
Our simplified system is composed of two mesoscopic and spherical particles, both of radius $a$,
 and a polymer chain \am{explicitly resolved by $N=60$ beads} of diameter $\sigma$ and interconnected by harmonic springs.
Here we choose the mesoscopic particle radius $a$ to be $5\sigma$.

We consider steric repulsion between all described particles through a WCA potential, which
 represents a purely repulsive interaction.
It is given by
\begin{equation}\label{WCA}
V_{WCA} \left(\frac{r'}{\sigma'}\right)=\left\{ {
\begin{array}{*{20}cl}
   4\varepsilon \left[ {\left( {\frac{r'}{\sigma' }} \right)^{\!-12}\!\!- \left( {\frac{r'}{\sigma' }}
 \right)^{\!-6}\!\!+ \frac{1}{4}} \right] & \mbox{for $r' \le r_c$}  \\
   \\
   0 & \mbox{otherwise}\\
\end{array}
} \right.
\end{equation}
where $r'$ is the distance between the particle centers, $\varepsilon$ denotes the energy scale of the potential, and $r_c=2^{1/6}\sigma'$ is the cut-off distance.
For any combination of large and small particles, $\sigma'$ is chosen as the sum of the radii of the respective particles, which is equivalent to the mean of their respective diameters.
In our simulations, we set $\varepsilon=10 k_BT$, where $k_B$ is the Boltzmann constant
 and $T$ is the temperature of the system.
Neighboring beads within the chain are bound by means of a harmonic potential
\begin{equation}\label{harmspring}
V_H(r') = \frac12 k (r'-r_0)^2,
\end{equation}
where we choose the force constant $k=10 k_BT/\sigma^2$. The equilibrium distance $r_0$ is set to match
 the cut-off  distance of the WCA-potential $r_0=r_c=2^{1/6}\sigma$.

\begin{figure}[h]
\centering
  \includegraphics[width=8.6cm]{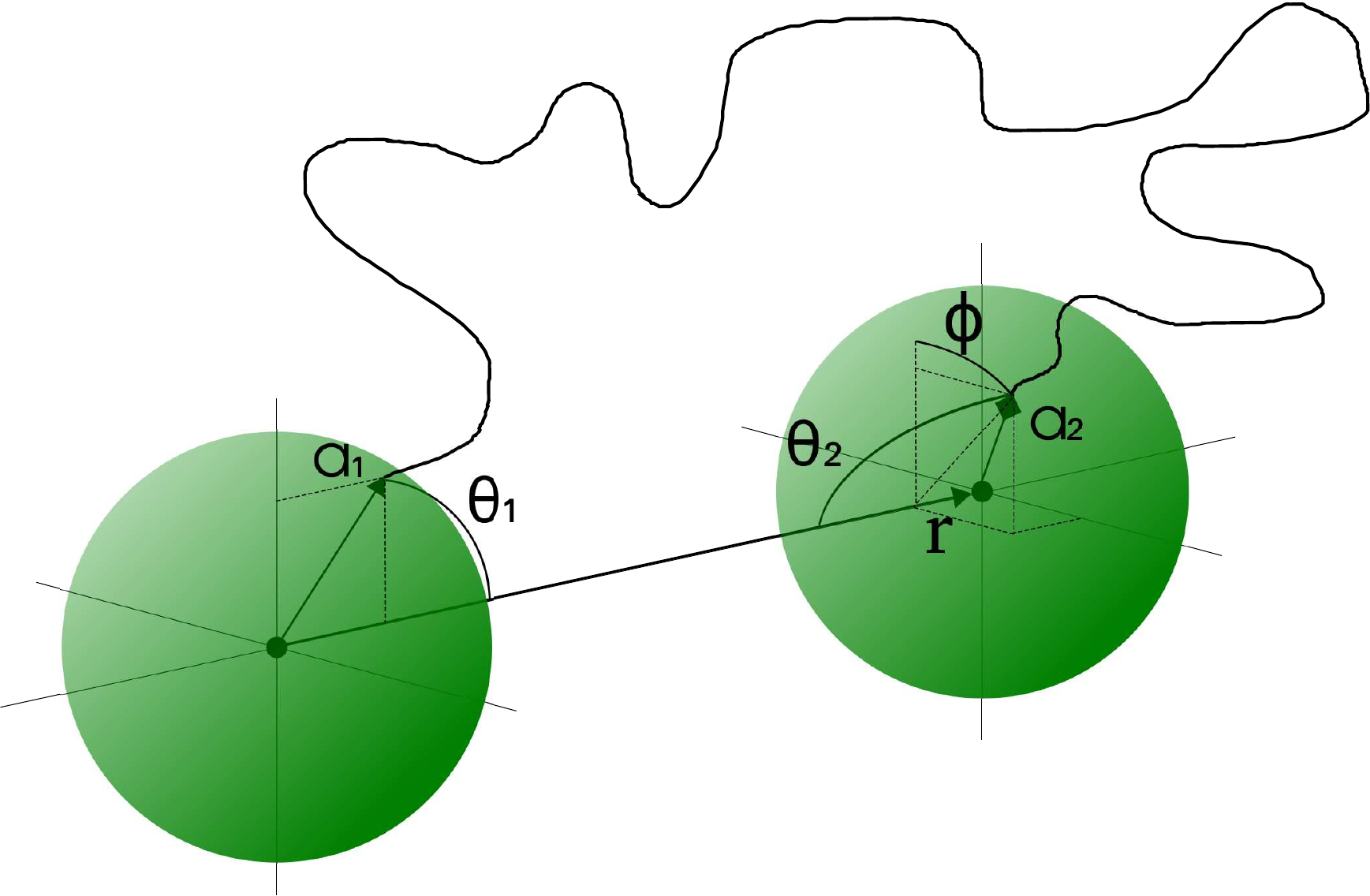}
  \caption{A simplified sketch of the geometry of the microscopic system.
 $\bm{r}$ is the vector connecting the centers of the mesoscopic particles.
 $\bm{a}_1$ and $\bm{a}_2$ identify the anchoring points of the polymer chain on the surfaces
 of the particles.
  $\theta_1$, $\theta_2$, and $\phi$ are the angles that represent the remaining relative rotational
 degrees of freedom of the system.}
  \label{scheme}
\end{figure}

The ends of the chains are bound via the same harmonic potential to binding sites placed below the surface
 of the mesoscopic particles, see figure \ref{scheme}.
These binding sites are rigidly connected to the mesoscopic particles and follow both, their translational
 and rotational motion.
Thus, when the magnetic particle moves or rotates, the binding site of the polymer chain
 has to follow, and vice versa.
The technical details for the virtual sites mechanism can be found in \cite{arnold13a}.
We identify the anchoring points of the polymer chain by the vectors $\bm{a}_1$ and $\bm{a}_2$, respectively
 (see figure \ref{scheme}).
Furthermore, the distance vector between the two mesoscopic particle centers is indicated
 as $\bm{r}$, with magnitude $r$.
The angles that the vectors $\bm{a}_1$ and $\bm{a}_2$ adopt with respect to the connecting
 vectors $\bm{r}$ and -$\bm{r}$ are denoted as $\theta_1$ and $\theta_2$, respectively.
$\theta_1$ and $\theta_2$ are zenithal angles and they can span the interval $[0,\pi]$.
Last, $\phi \in[-\pi,\pi[$ is the relative azimuthal angle between the projections of
 $\bm{a}_1$ and $\bm{a}_2$ on a plane perpendicular to $\bm{r}$.
It can be used to parametrize the relative torsion between the two particles around the $\bm{r}$ axis.
Let us for brevity introduce the vector $\gam=(r,\theta_1,\theta_2,\phi)$.
Therefore, the $\gam$-space, where our mesoscopic vector $\gam$ is defined, is given by $[0,+\infty[\times
[0,\pi]\times[0,\pi]\times[-\pi,\pi[$.

Through molecular dynamics simulations (thoroughly described in the next section) we find the probability
 density $p_c(\gam)$ of a certain configuration $\gam$ of the two mesoscopic particles.
\am{We normalize $p_c(\gam)$ such that $\int p_c(\gam) \d\gam=1$, with
$\d\gam = \sin(\theta_1) \sin(\theta_2) \d r \d\theta_1 \d\theta_2 \d\phi$
 the $\gam$-space volume element.
Moreover, we are working in the canonical ensemble.} 

\am{From statistical mechanics \cite{hansen2002effective}, we know that the probability density $p_c(\gam)$
 to find the system in a certain configuration $\gam$ is $p_c(\gam)=\exp[-\beta V_c(\gam)]/Z_c$,
 where $\beta=1/ k_BT$ and $Z_c$ is the partition function of the system. Shifting $V_c(\gam)$ by an appropriate constant, we can still reproduce $p_c(\gam)$ but simultaneously normalize $Z_c=1$.
Then, since $\ln(Z_c)=0$,
\begin{equation}\label{ec}
V_c(\gam) = -k_BT \ln\left[p_c(\gam)\right]
\end{equation}
represents an effective energy of the state $\gam$ of our mesoscopic two-particle system and corresponds to an effective \am{pair} potential on the mesoscopic level, see also \cite{harmandaris2007comparison}.
Through the normalization of $Z_c$, we set our reference free energy $F=-k_BT \ln(Z_c)$ equal to zero.} 

\section{Microscopic Simulation}\label{micro}
To obtain the probability distribution from which the mesoscopic \am{pair} potential is derived,
 we performed molecular dynamics simulations using the ESPResSo software \cite{limbach06a,arnold13a}.
Because entropic effects are important to capture the behavior of the polymer, the canonical
 ensemble is employed.
This is achieved using a Langevin thermostat, which adds random kicks as well as a velocity
 dependent friction force to the particles.
For the translational degrees of freedom of each particle, the equation of motion is then given by
\begin{equation}\label{motion}
m_p \dot{\bm{v}}(t) =-\zeta \bm{v}(t) +\bm{F}_r +\bm{F},
\end{equation}
where $m_p$ is the mass of a particle, $\bm{F}$ is the force due to the interaction with other particles,
 $\bm{F}_r$ denotes the random thermal noise, and $\zeta$ is the friction coefficient.

To maintain a temperature of $T$, according to the fluctuation-dissipation theorem, each random force
 component must have zero mean and variance $2 k_B T \zeta$.
Furthermore, random forces at different times are uncorrelated.
In order to track the orientation of the magnetic nanoparticles, rotational degrees of freedom also
 have to be taken into account.
This is achieved by means of a Langevin equation of motion similar to (\ref{motion}) where, however, mass,
 velocity, and force are replaced by inertial moment, angular velocity,
 and torque, respectively \cite{wang2002molecular}.

The time step for the integration using the Velocity-Verlet method \cite{frenkel02b} is $dt=0.01$.
To sample the probability distribution, we record the state of the simulation $\gam$ every ten time steps.
In order to obtain a smooth probability distribution over a wide range of parameters, 34 billion states have
 been sampled in total, by running many parallel instances of the simulation, summing
 up to about $10^4$ core hours of CPU time.

Finally, we find the probability distribution by sorting the results of our simulations into a histogram
 with 100 bins for each variable ($r$, $\theta_1$, $\theta_2$, and $\phi$).
When a 64-bit unsigned integer is used as data type, this leads to a memory footprint of 800 Mb.
Hence, the complete histogram can be held in memory on a current computer.
If the resulting numerical version of the effective \am{pair} potential as defined in (\ref{ec}) is to be used
 in a simulation, a smoothing procedure should be employed, especially in parts of the configuration
 space with a very low probability density.
One approach here might be hierarchical basis sets.

In our simulations, we chose the thermal energy $k_B T$ as well as the mass $m_p$, the friction
 coefficient $\zeta$, and the corresponding rotational quantities to be unity.
We measure all lengths in units of $\sigma$ and the energies in multiples of $k_BT$.

\section{Description of the Probability Density}\label{stat}
We now consider some aspects of the probability density resulting from the microscopic
 simulation described in section \ref{micro}.
The entropic role of the microscopic degrees of freedom is considered by assigning to every
 configuration $\gam$ a certain probability, given by the number of times it was encountered
 in the simulation divided by the total number of recorded states.
As explained before, $\gam=(r,\theta_1,\theta_2,\phi)$ corresponds to the set of variables
 that we use to describe the state of the system on the mesoscopic level.

In calculating the probability density from the microscopic simulations, we must remember
 the normalizing condition
\begin{equation}\label{norm}
\int p_c(\gam) \sin(\theta_1) \sin(\theta_2) \d r \d\theta_1 \d\theta_2 \d\phi = 1,
\end{equation}
where $r$ is integrated over $[0,+\infty[$, $\theta_1$ and $\theta_2$ over $[0,\pi]$, and $\phi$ over
 $[-\pi,\pi[$.
Therefore, to obtain the probability density, we have to properly divide the data acquired from the simulations 
by the factor $\sin(\theta_1) \sin(\theta_2)$.

It is useful to introduce here the average of a quantity over $p_c$, defined as
 $\meanc{\cdot} = \int \cdot \ p_c(\gam) \d\gam$.
We can therefore calculate the average value of $\gam$, $\overline{\gam}=\meanc{\gam}$, and the covariance matrix
 $\Sigma^c_{\alpha\beta}=\meanc{ \left(\alpha - \overline{\alpha} \right)\left(\beta -\overline{\beta}  \right) }$,
 for $\alpha,\beta = r, \theta_1,\theta_2,\phi$.
We find  $\overline{\gam}\simeq(20.20\sigma, 0.36\pi, 0.36\pi, 0)$.
It is more practical to discuss the system in terms of correlation than in terms of covariance.
Correlation is defined as ${\varrho}^c_{\alpha\beta}=\Sigma^c_{\alpha\beta}/\sqrt{\Sigma^c_{\alpha\alpha}\Sigma^c_{\beta\beta}}$ (no summation rule in this expression), is dimensionless, and ${\varrho}^c_{\alpha\beta}\in[-1,1]$.
Here, we obtain
\widetext
\begin{align}
{\bm{\varrho}^c}\simeq
 \begin{pmatrix}
  1 & -0.341\pm0.028 & -0.356\pm0.029 & 0.083\pm0.022 \\
  -0.341\pm0.028 & 1 & -0.006\pm0.006 & -0.083\pm0.010 \\
  -0.356\pm0.029 & -0.006\pm0.006 & 1 & -0.083\pm0.011 \\
  0.083\pm0.022 & -0.083\pm0.010 & -0.083\pm0.011 & 1
 \end{pmatrix},
\label{corrc}
\end{align}
\endwidetext
\noindent where lines and columns refer to $ r, \theta_1,\theta_2,\phi$ in this order.
The diagonal elements are unity by definition, because each variable is perfectly correlated with itself.
\am{The errors follow from the unavoidable discretization during the statistical sampling procedure in the simulations, where the results have to be recorded in discretized histograms of finite bin size.}

We find a strong anticorrelation between $r$ and $\theta_{1,2}$, meaning that when the distance between the mesoscopic particles changes they tend to rotate.
We will address in detail the background of this behavior in section \ref{wrap} in the form of the wrapping of the polymer chain around the mesoscopic particles.
\am{For angles $\theta_1$ and $\theta_2$ different from $0$ and $\pi$ it is clear that such a wrapping and corresponding distance changes are likewise induced by modifying the relative torsion of angle $\phi$.
These coupling effects are partly reflected by the remaining non-vanishing correlations which, however, are weaker than the correlations between $r$ and $\theta_{1}$, $\theta_{2}$ by at least a factor $4$.
The correlations between  $\theta_1$ and $\theta_2$ are very weak and, within the statistical errors, may in fact even vanish.
Within the statistical errors, the magnitudes of the correlations $\varrho^c_{r,\theta_1}$ and  $\varrho^c_{r,\theta_2}$ as well as $\varrho^c_{\theta_1,\phi}$ and  $\varrho^c_{\theta_2,\phi}$ agree well with each other, respectively, which reflects the symmetry of the system.}
\am{Finally, we performed additional microscopic simulations for different sizes of mesoscopic particles and different chain lengths of the connecting polymer. As a general trend, we found that the correlations tend to decrease in magnitude for smaller mesoscopic particles and for longer polymer chains (see appendix \ref{app2}).
}

\begin{figure}[]
\centering
  \includegraphics[width=8.6cm]{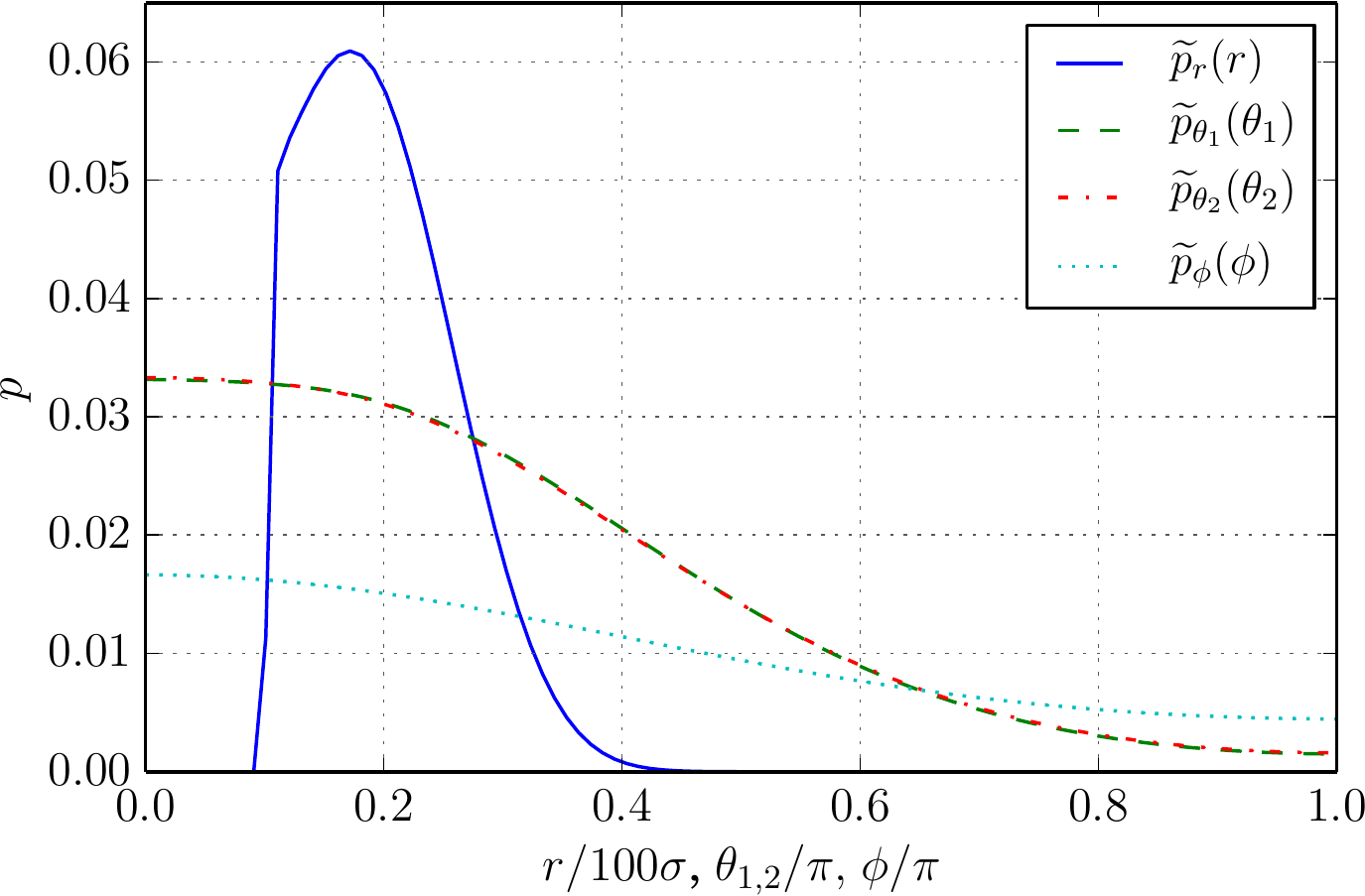}
  \caption{Marginal probability densities for the single variables.
The abscissa has been rescaled: $r$ ranges from $0$ to $100\sigma$, whereas the angles
 range from $0$ to $\pi$.
$\ptil_{\theta_1}(\theta_1)$ and $\ptil_{\theta_2}(\theta_2)$ are practically indistinguishable
 due to the symmetry of the set-up.
Due to the symmetry of the system under the transformation $\phi \rightarrow -\phi$,
 here we only  plot $\ptil_\phi(\phi)$ from $0$ to $\pi$.
The maxima of the single-variable densities are located at $r=17\sigma$ and $\theta_{1,2}=\phi=0$,
 respectively.
The maxima of $\ptil_{\theta_1}(\theta_1)$ and $\ptil_{\theta_2}(\theta_2)$ at $\theta_{1,2}=0$
 are found from the microscopic simulations after the $\gam$-space normalization
 contained in (\ref{norm}) has been taken into account.}
  \label{figpartial}
\end{figure}

As a further step in the analysis of $p_c$, we can determine the marginal probability density $\ptil_\alpha(\alpha)$
 for one of the four mesoscopic parameters $\alpha = r,\theta_1,\theta_2,\phi$ integrating out the other three,
 for instance, $\ptil_{\theta_1}(\theta_1)=\int p_c(\gam) \sin(\theta_2) \d r \d\theta_2 \d\phi$.
This is the total probability density for the variable $\theta_1$ to assume a certain value, regardless
 of the others.
Calculations for $\ptil_{r}(r)$, $\ptil_{\theta_2}(\theta_2)$, and $\ptil_\phi(\phi)$
 are analogously performed by integrating out all respective other variables (see figure \ref{figpartial}).
Of course, $\ptil_\alpha(\alpha)$ is still a normalized probability density since
 $\int \widetilde{p}_\alpha(\alpha) \d\alpha= \int p_c(\gam) \d\gam = 1$, where
 we indicate $\d\alpha = \sin(\theta_i)\d\theta_i$ for $\alpha = \theta_i$ ($i=1,2$)
 and $\d\alpha=\d r,\d\phi$ for $\alpha=r,\phi$.
Following (\ref{ec}), we also introduce the one-variable effective \am{pair} potentials
\begin{equation}\label{1veffp}
 \vtil_\alpha(\alpha)= -k_BT\ln[\ptil_\alpha(\alpha)]
\end{equation}
 that are associated with the corresponding single-variable marginal probability density.

In figure \ref{figpartial} most of the probability density $\ptil_r(r)$ for the interparticle
 distance is contained between $r=10\sigma$ and $r=50\sigma$, with a single maximum at $r=17\sigma$.
Moreover, the steep increase in $\ptil_r(r)$ at $r=11\sigma$ is to be attributed to the WCA
 steric repulsion, since at that distance the two particles are in contact.
From figure \ref{figpartial} we find that the maximum of $\ptil_{\theta_i}(\theta_i)$ (with $i=1,2$)
 is located at $\theta_i=0$.
The highest probability density for $\theta_{1,2}=0$ is obtained from the microscopic simulations
 by taking into account the $\gam$-space normalization following
 from the use of spherical coordinates in (\ref{norm}), see also \cite{harmandaris2007comparison}.
Moreover, $\ptil_{\phi}(\phi)$ shows a maximum for $\phi=0$, indicating that, as
 expected, the system does not tend to spontaneously twist around the connecting axis in the absence of further interactions.
The presence of a maximum at $\phi=0$ confirms that $\ptil_{\phi}$ is an even function of $\phi$ invariant under
 the transformation $\phi \rightarrow -\phi$, as expected from the symmetry of the set-up.

\section{Wrapping Effect}\label{wrap}
Before developing an approximate analytical expression for the effective \am{pair} potential between the mesoscopic particles,
 it is helpful to examine in detail the results of the molecular dynamics simulations.
In a magnetic gel in which mesoscopic magnetic particles act as cross-linkers \cite{messing2011cobalt,frickel2011magneto},
 two driving mechanisms for a deformation in an external magnetic field are possible.
First, in any magnetic gel, the magnetic interactions between the mesoscopic particles lead to attractions
 and repulsions between them, which directly implies deformations of the intermediate polymer chains.
As we would like to examine this mechanism separately, the magnetic interaction was not included explicitly
 in the simulations described in section \ref{micro}.
Rather, it will be considered later in section \ref{magneff}.
Second, due to the anchoring of the polymer chains on the surfaces, rotations of the mesoscopic particles
 are transmitted to chain deformations.

\begin{figure}
\includegraphics[width=8.6cm]{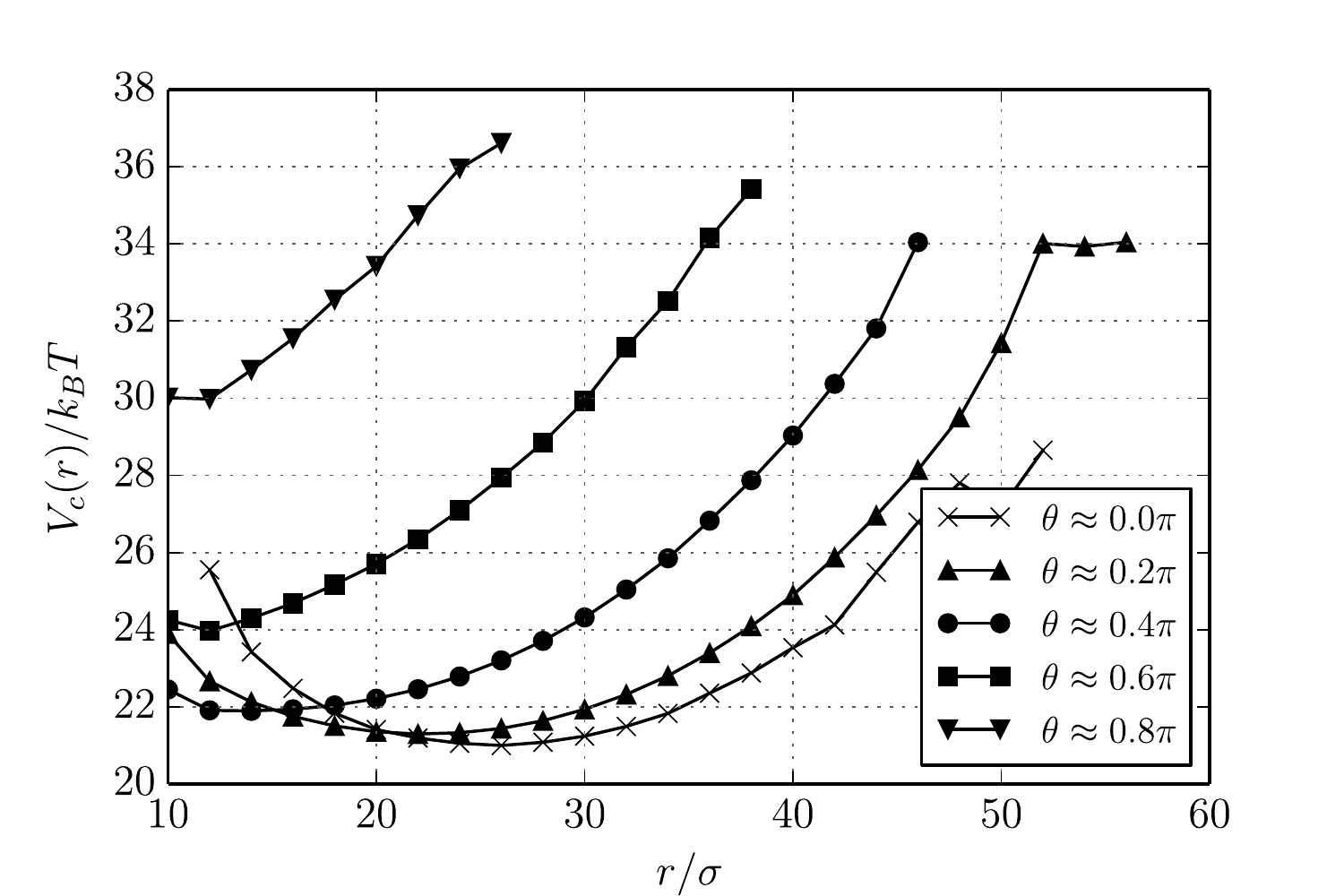}
\caption{
Plot of the effective mesoscopic \am{pair} potential $V_c(r)$ for a situation in which both particles are rotated by the same amount in the angles $\theta_{1}=\theta_{2}=\theta$ at a torsional angle of $ \phi=0$.
It can be seen that the more the particles are rotated out of their equilibrium position,
 the closer they will approach each other because the minimum of the effective \am{pair} potential
 shifts to smaller separation distances $r$.
The irregularities in the effective \am{pair} potential for high values of $r$ are attributed
 to the naturally low sampling of those low-probability configurations.}
 \label{fig:u-of-r}
\end{figure}

It has been shown in model II of \cite{weeber2012deformation} that the second mechanism on
 its own can lead to a deformation of such a gel in an external magnetic field: if mesoscopic magnetic
 particles are forced to rotate to align with the field, the polymers attached to their surfaces have to follow.
The resulting wrapping of the polymer chains around the particles leads to a shrinking of the gel.
Transferring this to our model system, it could imply an external magnetic field rotating the particles
 to a state in which the angles $\theta_1$ and $\theta_2$ are non-zero. 
The effect of induced particle rotations on the interparticle distance is illustrated in figure \ref{fig:u-of-r},
 where the effective \am{pair} potential $V_c(r)$ is plotted for the case of $ \phi=0$ and various values
 of $\theta_1=\theta_2=\theta$.
I.e., both particles are rotated by the same angle $\theta$ with respect to the connecting
 vector $\pm \bm{r}$, respectively.
It can be seen that the more the particles are rotated, the more the minimum of the effective \am{pair} potential is shifted
 towards closer interparticle distances corresponding to smaller separation distances $r$.

\begin{figure}
\includegraphics[width=8.6cm]{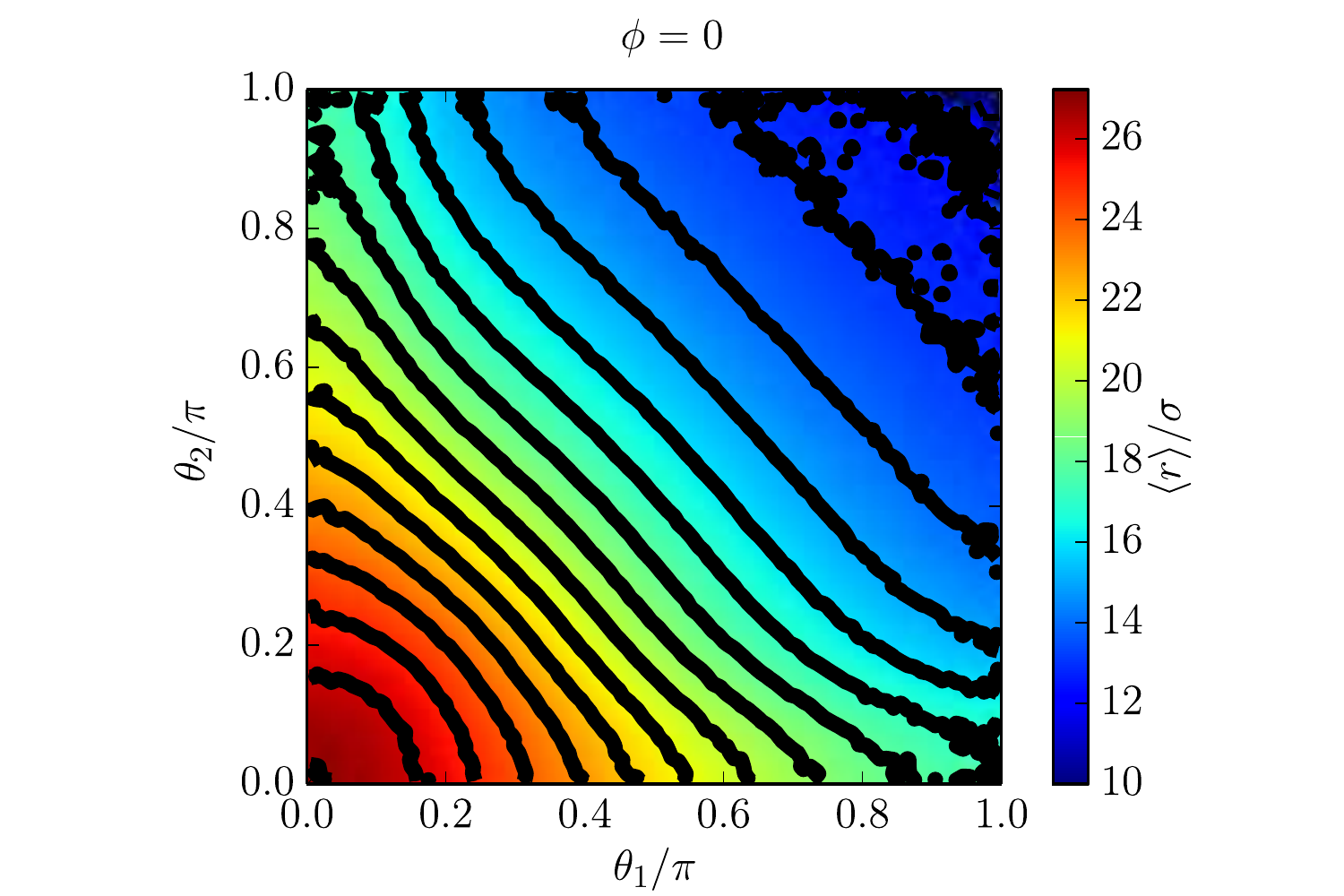}\hfill
\includegraphics[width=8.6cm]{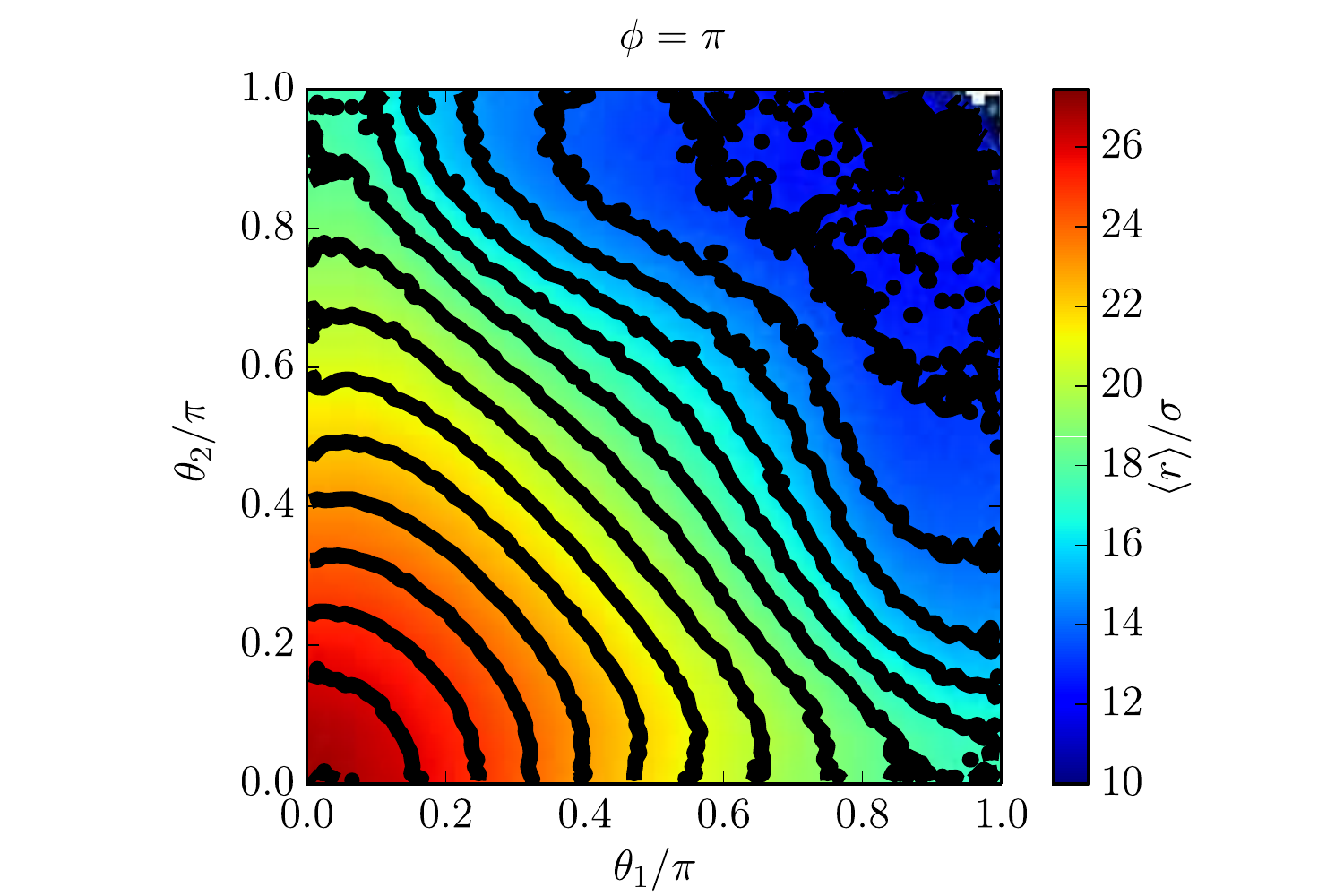}
\caption{
Average distance between the two mesoscopic particles versus the angles $\theta_1$ and $\theta_2$
 shown as a color and contour map for the two cases $\phi=0$ (top) and $ \phi = \pi$ (bottom).
It can be seen that even by rotating only one of the two particles, the average distance can be reduced considerably.
However, the maximum reduction is observed when both particles are rotated.
The variation of the torsion angle $\phi$ does not change the main trend but has a moderate influence
 for intermediate rotations.}
 \label{fig:avg-dist}
\end{figure}

To get a more detailed picture, we also consider independent rotations of the two mesoscopic particles.
In Fig. \ref{fig:avg-dist}, the average distance between the particles is depicted in a contour plot
 as a function of the angles $\theta_1$ and $\theta_2$.
Images are shown for torsion angles of $ \phi =0$ and $ \phi = \pi$.
It can be seen that quite a decrease in the average distance can already be achieved by rotating only one particle.
However, very strong reduction in interparticle distance can only occur when both particles are rotated.
The torsion angle $ \phi$ does not alter the general trend of reduction of the average distance
 when the particles are rotated.
However, the resulting numbers vary to a certain degree.

\section{One-Variable Effective \am{Pair} Potentials}\label{margpot}
We now introduce some mesoscopic analytical expressions to model the effective \am{pair} potentials
 $\vtil_\alpha(\alpha)$ introduced in (\ref{1veffp}).
We will determine functional forms and parameters that can be used to model strain and torsion energies.
Here, we use the term ''strain`` to denote variations of $r$, whereas the term ''torsion``
 is used to describe changes in the angles $\theta_1$ and $\theta_2$ or $\phi$, depending
 on the initial orientation of the spheres and relative rotations.
The quality of the analytical model expressions will be analyzed by fitting to the corresponding
 results from the microscopic simulations.

\begin{figure}[]
\centering
  \includegraphics[width=8.6cm]{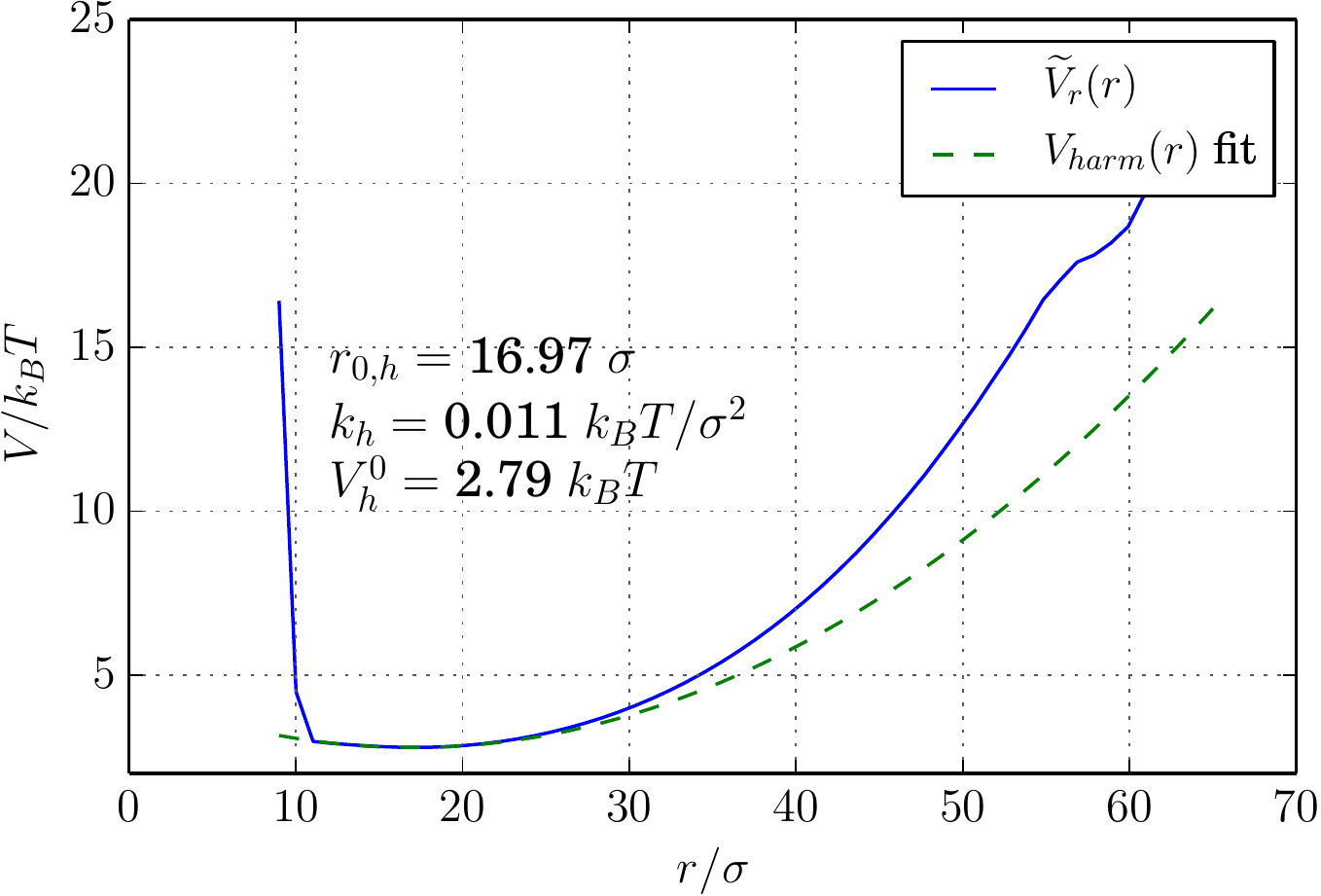}
  \caption{Effective \am{pair} potential $\vtil_{r}(r)$ obtained from the microscopic simulation data
 and fit using a simple expression $V_{harm}(r)$ as in (\ref{harm}).
The fit was made in the interval $[11\sigma,27\sigma]$.
In this way, the values for the mesoscopic model parameters $r_{0,h},k_h,V_h^0$ are determined.}
  \label{fiter}
\end{figure}

First, we turn to the energetic contributions arising from changes in the interparticle distance $r$.
The corresponding effective energy $\vtil_r(r)$ obtained from the microscopic data is plotted in figure \ref{fiter}.
It can be seen that there are essentially two regimes: up to $r\simeq 11\sigma$
 the WCA repulsion between the two mesoscopic particles dominates, whereas, for $r > 11\sigma$,
 $\vtil_r(r)$ shows a smooth behavior and a single minimum arising from the entropic contribution of the polymer chain.
Moreover, at $r \gtrsim 55\sigma$, $\vtil_r(r)$ shows an irregular, non-smooth behavior.
This is attributed to the low sampling rate of this extremely stretched configuration, which
 has a very low probability to occur in the microscopic simulations (see figure \ref{figpartial}).
As a first approximation, it is natural to reproduce $\vtil_r(r)$ by a harmonic expansion for $r>11\sigma$,
\begin{equation}\label{harm}
V_{harm}(r) =  V_h^0 + \frac{k_h}{2}{(r-r_{0,h})}^2.
\end{equation}
We derive the mesoscopic parameters $V_h^0,k_h,r_{0,h}$ by fitting $V_{harm}(r)$ in a neighborhood
 of the minimum to the data $\vtil_r(r)$ obtained from microscopic simulations.
In figure \ref{fiter} the resulting parameters and the two curves are shown.

\begin{figure}[]
\centering
  \includegraphics[width=8.6cm]{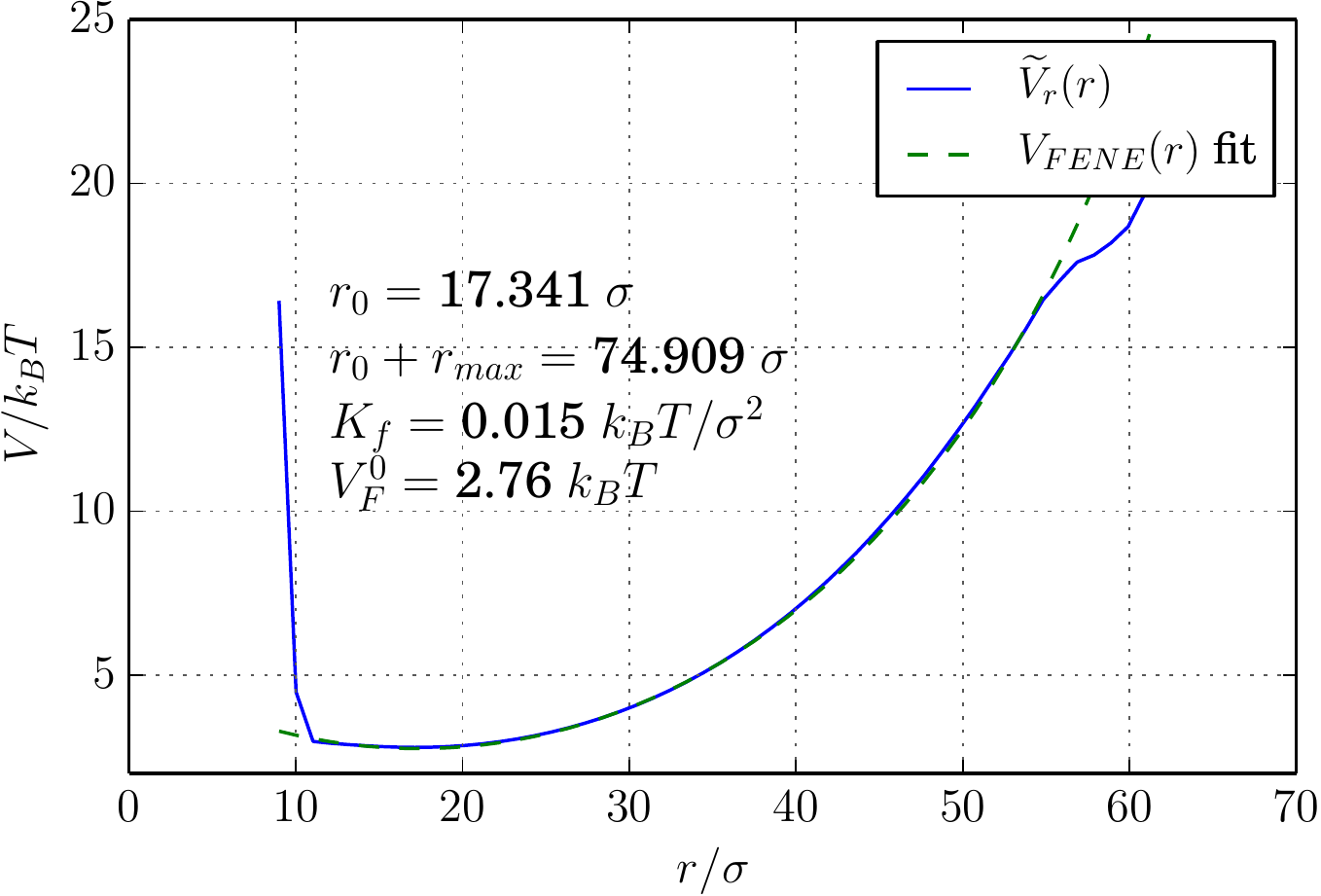}
  \caption{Effective \am{pair} potential $\vtil_{r}(r)$ obtained from the microscopic simulation data
 and fit using ${V}_{FENE}(r)$ from (\ref{fene}) leading to the parameter values as listed in the plot.
The fit was made in the interval $[11\sigma, 52\sigma]$.
In this way, the values for the mesoscopic model parameters $r^0,r_{max},K_f,V_F^0$ are determined.}
  \label{fitfene}
\end{figure}

Moreover, we have compared $\vtil_r(r)$ with the following expression for a finitely
 extensible non-linear elastic potential (FENE) \cite{warner1972kinetic,sanchez2013effects},
\begin{equation}\label{fene}
V_{FENE}(r)= V_{F}^0 -\frac{K_f\ r^2_{max}}{2} \ln\left[1- {\left( \frac{r-r_0}{r_{max}} \right)}^2 \right].
\end{equation}
It takes into account that the chain cannot extend beyond a maximal length,
 since $V_{FENE}(r)$ diverges when $r \rightarrow r_0+r_{max}$.
We see that $d^2 V_{FENE}(r)/ {dr}^2 = K_f$ for $r=r_0$ and therefore $K_f$ represents
 the elastic constant in a harmonic expansion of this non-linear potential.
As for the harmonic expression, we fit ${V}_{FENE}(r)$ to our microscopic data and thus derive the
 mesoscopic model parameters $K_f$, $r_0$, $r_{max}$, and $V_{F}^0$ as displayed in figure \ref{fitfene}.
The agreement between the resulting expression for ${V}_{FENE}(r)$ and $\vtil_r(r)$ in the regime
 $r\gtrsim 11 \sigma$ is excellent, especially for the branch of the curve right to the minimum.
According to the fit, the maximum extension of the chain occurs for $r=r_0+r_{max}\simeq 75 \sigma$.
In fact, since the radius of the mesoscopic particles is $5\sigma$ and each of the $60$ beads making up
 the polymer chain has diameter $\sigma$, at $r=70\sigma$ the polymer chain is completely stretched.
A further elongation is of course possible due to the harmonic inter-bead interaction
 and this justifies the result of $\sim 75 \sigma$ for the maximal extension.

Last, we compare the elastic constants $k_h$ and $K_f$ obtained from the harmonic and FENE approximation,
 respectively, as listed in figures \ref{fiter} and \ref{fitfene}.
The resulting values of $0.011k_BT/\sigma^2$ and $0.015k_BT/\sigma^2$ are in good agreement with each other.

\begin{figure}[]
\centering
  \includegraphics[width=8.6cm]{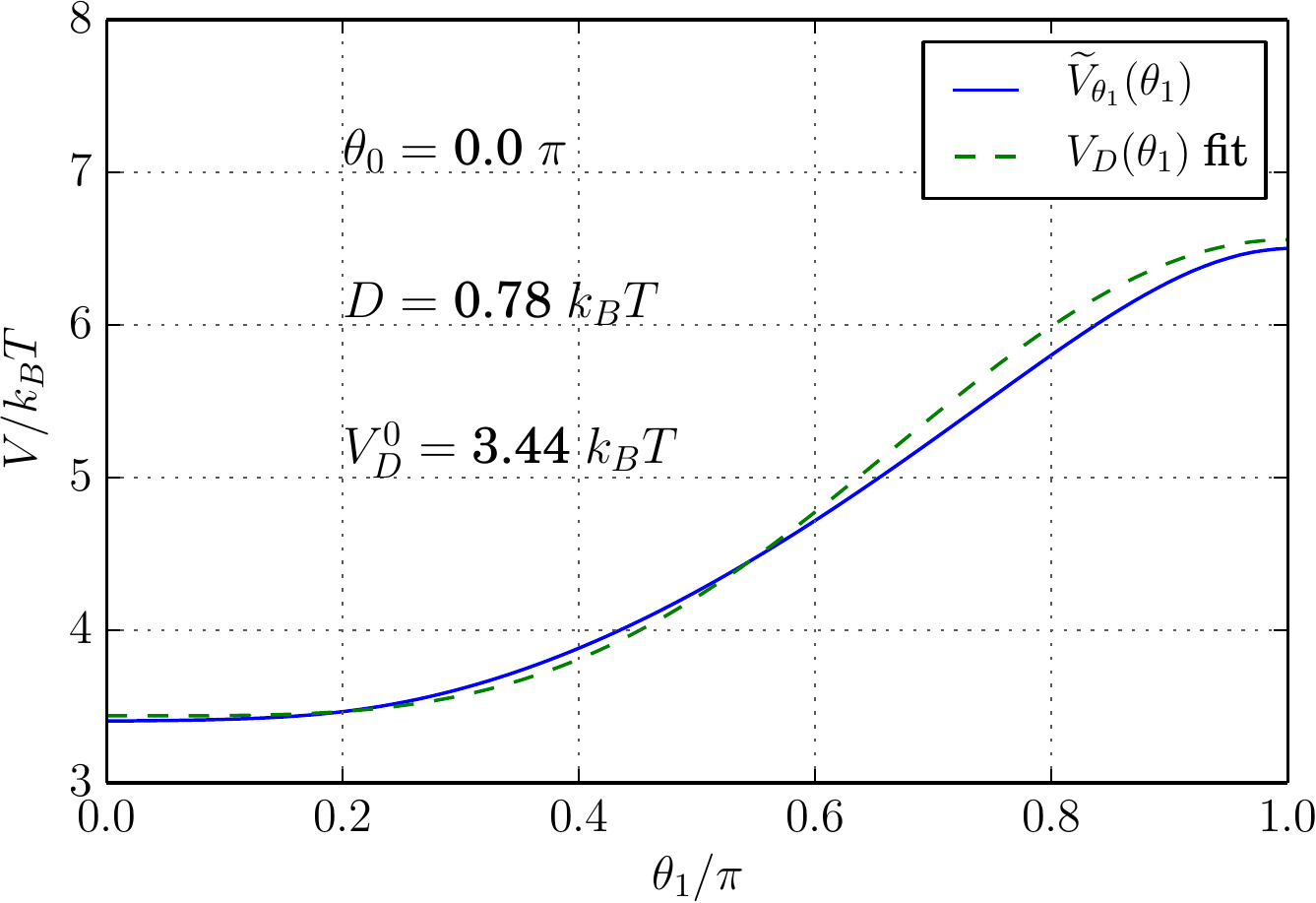}
  \caption{Effective \am{pair} potential $\vtil_{\theta_1}(\theta_1)$ calculated from the
 microscopic simulation data and fit using the phenomenological mesoscopic expression
 $V_D(\theta_1)$ from (\ref{ed}).
 Resulting values for the mesoscopic model parameters are listed in the plot.
 The fit was made on the interval $[0,\pi]$.}
  \label{fited}
\end{figure}

To find a mesoscopic model expansion for the effective \am{pair} potential $\vtil_{\theta_1}(\theta_1)$
[$\vtil_{\theta_2}(\theta_2)$ has a very similar behavior], we compare it with
 the phenomenological expression,
\begin{equation}\label{ed}
V_D(\theta_1) = V_D^0 + D {\left[ \cos(\theta_1) - \cos(\theta_0) \right]}^2
\end{equation}
 introduced in (3) of \cite{annunziata2013hardening}.
As before, we can derive the mesoscopic parameters $V_D^0$, $D$, and $\theta_0$  by fitting
 ${V}_{D}(\theta_1)$ to the microscopic data represented by $\vtil_{\theta_1}(\theta_1)$.
The resulting parameters and the comparison between the two curves are shown in figure \ref{fited}.
Although (\ref{ed}) does not perfectly reproduce the one-variable \am{pair} potential
 $\vtil_{\theta_1}(\theta_1)$, it appears as a reasonable approximation in the neighborhood of the
 minimum energy.
Moreover, the rather flat behavior of $\vtil_{\theta_1}(\theta_1)$ for small $\theta_1$
 is well represented by $V_D(\theta_1)$, which is at leading order proportional to ${\theta_1}^4$.

\begin{figure}[]
\centering
  \includegraphics[width=8.6cm]{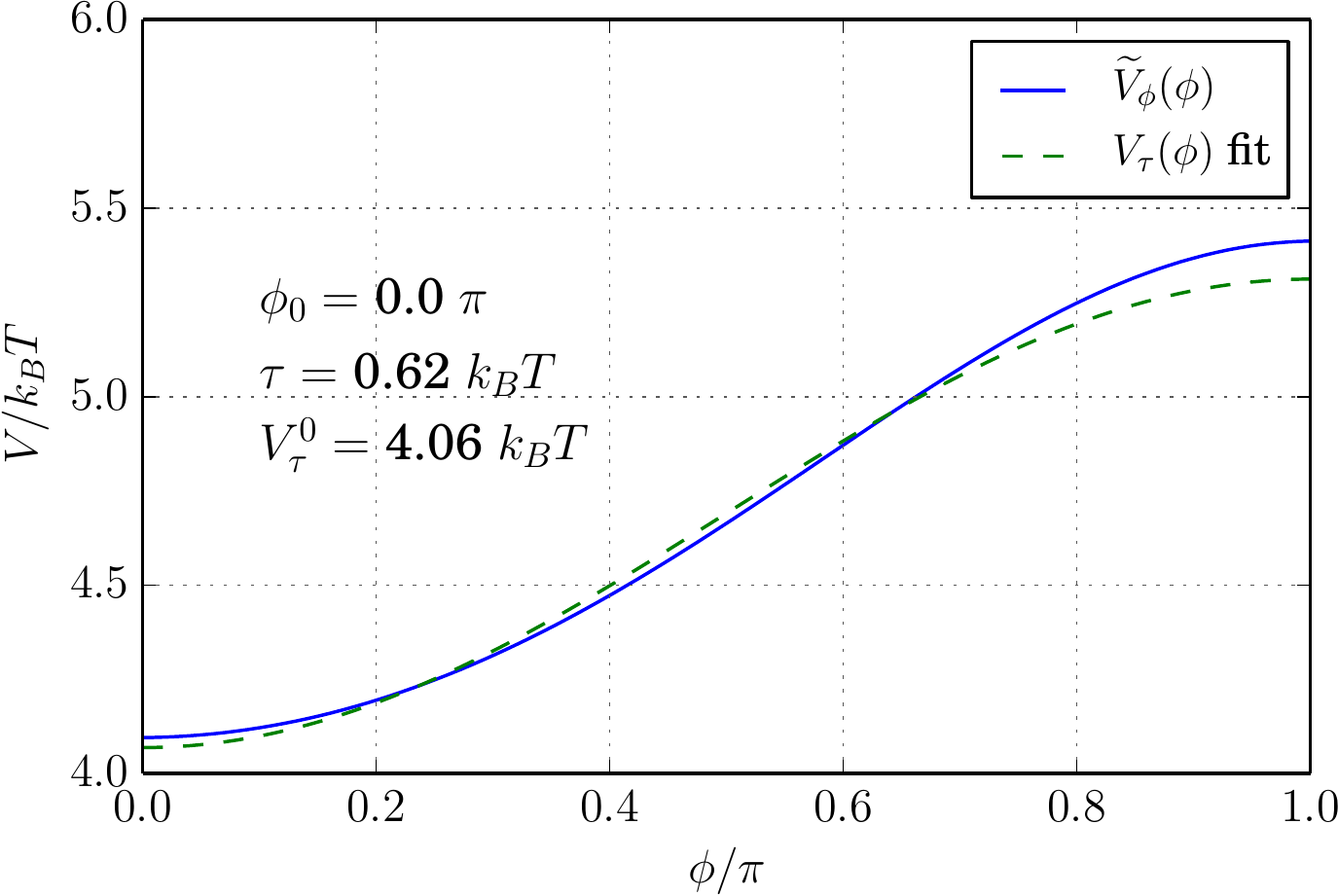}
  \caption{Effective \am{pair} potential $\vtil_\phi(\phi)$ calculated from the
 microscopic simulation data and fit using ${V}_\tau(\phi)$ from (\ref{etau}).
Resulting values for the mesoscopic model parameters are listed in the plot.
The fit was made in the interval $[0,0.8\pi]$.}
  \label{fittp}
\end{figure}

Finally, we want to find a mesoscopic expression to reproduce the effective \am{pair} potential
 $\vtil_{\phi}(\phi)$ obtained from the microscopic data due to relative torsional
 rotations between the two particles around the connecting vector $\bm{r}$.
We fit the microscopic data $\vtil_{\phi}(\phi)$ around the minimum with the expression
\begin{equation}\label{etau}
{V}_\tau(\phi) = {V}_\tau^{0}  +\tau {\left[ \cos(\phi_0) - \cos(\phi) \right]}.
\end{equation}
It leads to the parameters and fit depicted in figure \ref{fittp} and is quadratic
 at lowest order in $\phi$, ${V}_\tau(\phi) \simeq  {V}_\tau^0  +\tau \phi^2$.
The parameter $\phi_0$ is redundant and can be absorbed into ${V}_\tau^0$, but we
 leave it for reasons of comparison to the following (\ref{etau'}).

\begin{figure}[]
\centering
  \includegraphics[width=8.6cm]{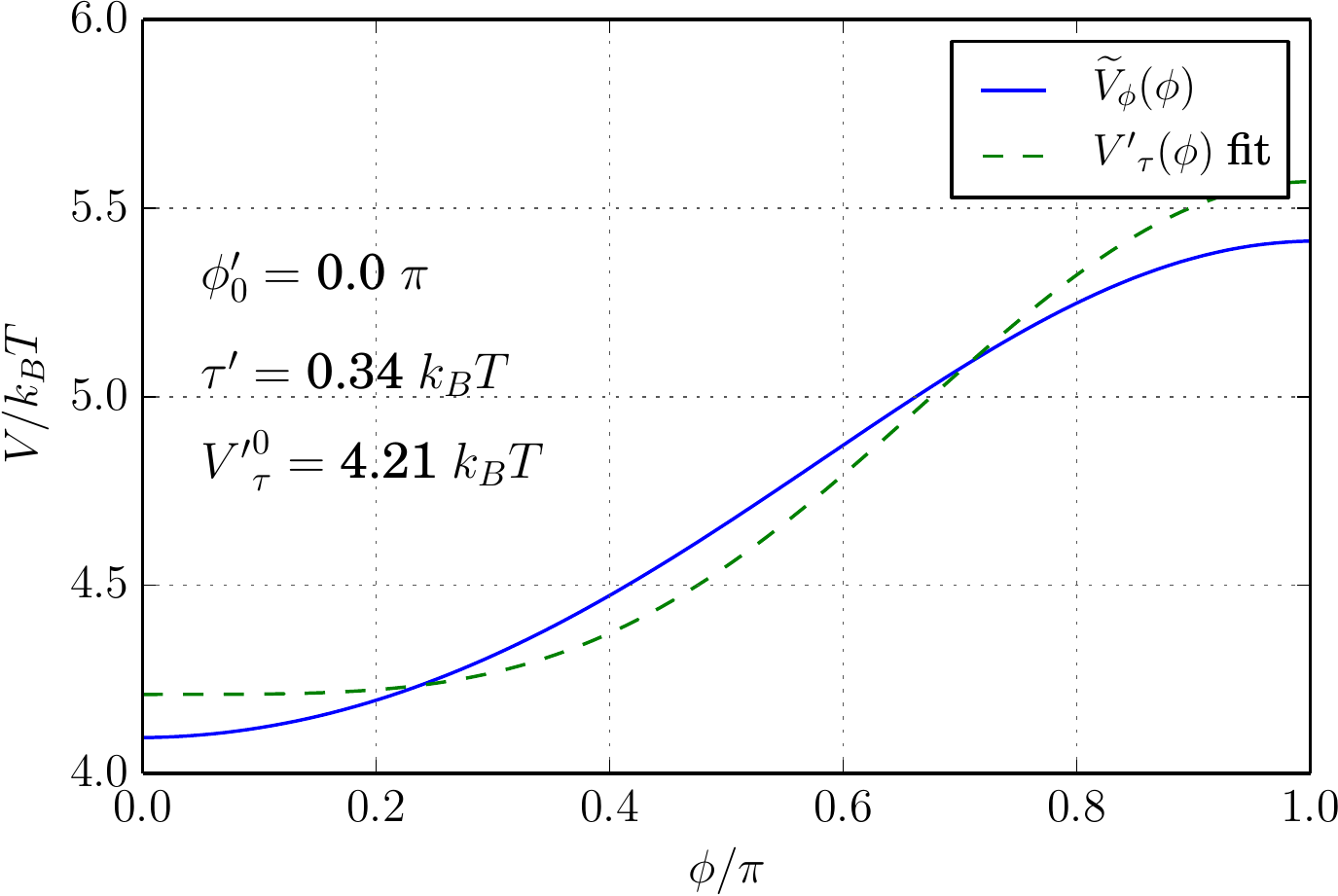}
  \caption{Effective \am{pair} potential $\vtil_\phi(\phi)$ calculated from the
 microscopic simulation data and fit using ${V'}_\tau(\phi)$ from (\ref{etau'}).
Resulting values for the mesoscopic model parameters are listed in the plot.
The fit was made in the interval $[0,0.8\pi]$.}
  \label{fitt1}
\end{figure}

It is interesting to compare $\vtil_\phi(\phi)$ with the expression taken from (4)
 of \cite{annunziata2013hardening},
\begin{equation}\label{etau'}
{V'}_\tau(\phi) = {V'}_\tau^0  +\tau' {\left[ \cos(\phi_0') - \cos(\phi)  \right]}^2.
\end{equation}
The discrepancy between the two curves shown in figure \ref{fitt1} is obvious.
The reason becomes clear when we expand (\ref{etau'}) to lowest order in
 $\phi$, ${V'}_\tau(\phi) \simeq  {V'}_\tau^0  +\tau' \phi^4$.
This expression is quartic in the torsion angle $\phi$, leading to the comparatively flat
behavior in the region around the minimum.
The last two comparisons suggest that the analytical form of the mesoscopic model \am{pair} potential
 as a function of the azimuthal angle, which acts as a torsion, can be optimized by using
 a quadratic form as in (\ref{etau}) instead of the quartic one implied by (\ref{etau'})
 at leading order around the minimum.

\am{
Finally, we performed additional microscopic simulations to estimate how the microscopic system parameters affect the mesoscopic model parameters. In particular, the influence of the mesoscopic particle radius $a$ and the number of beads $N$ of the connecting polymer chain was investigated.
As a trend, we found that the single-variable potentials tend to become stiffer for shorter chains and for bigger mesoscopic particles (see appendix \ref{app2}).
Moreover, we observe that the fit of the functional forms in expressions (\ref{fene})--(\ref{etau}) with the simulation data further improves for increasing chain length and decreasing size of the mesoscopic particles \cite{suppl}.
}

So far, we have discussed mesoscopic analytical expressions to approximate the one-variable
 effective \am{pair} potentials $\vtil_\alpha(\alpha)$.
In the following section, using a numerical fitting procedure, we will determine
 harmonic coupling terms that take account of the correlations between the mesoscopic variables.
In this way, we will further develop and improve our approximation of the probability
 density $p_c(\gam)$ from the microscopic simulations.

\section{Building a coupled effective \am{pair} potential}\label{effpot}
Our goal is to describe the effective interaction between the two mesoscopic particles, coarse-graining
 the role of the connecting polymer chain.
The entropic nature of the polymeric interactions provides a direct route to average out the microscopic
 degrees of freedom and thus build an effective scale-bridged model.
A natural way to proceed would be to find an analytical approximation for $p_c(\gam)$ and thus
 derive an analytical expression for the mesoscopic \am{pair} potentials in the spirit of (\ref{ec}).

As a first approximation we may describe $p_c$ as the simple product of the one-variable
 probability densities, $\prod_\alpha \ptil_\alpha(\alpha)$.
We obtain
\begin{align*}\label{punc}
 {p'}_{app}(\gam)=\exp\bigl\{-\beta &\bigl[ V_{WCA}(r/2a) + V_{FENE}(r) \\ 
&+ V_D(\theta_1) + V_D(\theta_2) +V_\tau(\phi)\bigr]\bigr\},
\addtocounter{equation}{1}\tag{\theequation}
\end{align*}
when we use (\ref{1veffp}).
Analytical approximations for the \am{pair} potentials $\vtil_\alpha(\alpha)$ were derived in (\ref{fene})-(\ref{etau})
 in section \ref{margpot} together with the fitting parameters listed in figures \ref{fitfene}-\ref{fittp}.
Care must be taken for the term $\vtil_r(r)$, which has to be substituted by
 $V_{WCA}(r/2a) + V_{FENE}(r)$ to take account of the steric repulsion between
 the mesoscopic particles, see (\ref{WCA}).
(\ref{punc}) correctly describes some aspects of the system behavior.
For instance, the steep variation due to the WCA potential between the mesoscopic particles is well represented
 in this approximation.
However, this description would lead to a distribution with vanishing correlations between the mesoscopic variables.
That corresponds to assuming them independent of each other, which omits some important physical aspects,
 see the wrapping effect in section \ref{wrap}.

To make a step forward and take account of correlations we multiply correction terms to the previous
 factorized approximation in (\ref{punc}), obtaining the expression 
\begin{equation}\label{papp}
 \papp(\gam) \propto {p'}_{app}(\gam) \times
 \exp\left[ - \left(\gam-\bm{\xi}\right) \cdot \bm{\Xi} \cdot \left(\gam-\bm{\xi}\right) \right],
\end{equation}
where the elements of $\bm{\xi}$, a $4$-components vector, and $\bm{\Xi}$, a $4\times 4$ symmetric matrix,
 are free parameters chosen to match the original data.
There are at least two possible numerical approaches to find the best $\bm{\xi}$ and $\bm{\Xi}$ values.
On the one hand, we can simply fit the original probability density (e.g.~minimize the squared difference
 between $p_c$ and $\papp$).
On the other hand, we can follow a moment-matching approach, looking for a $\papp$ that has
 a correlation matrix as close as possible to the original one.
We performed a mixed strategy by fitting the expression in (\ref{papp}) to the simulation
 data $p_c(\gam)$, using as a criterion for best initialization of the fit an outcome that as
 close as possible matches the correlations directly calculated from the simulation data $p_c(\gam)$.
This fit was performed by minimizing the squared difference between $p_c(\gam)$
 and $\papp(\gam)$, using the Nelder-Mead algorithm \cite{nelder1965simplex}
 provided by the SciPy library \cite{jones2001scipy}.
In this procedure, normalization of the approximated probability density as in (\ref{norm}) is enforced.
The best set of parameters $\bm{\xi}$ and $\bm{\Xi}$ found is
\begin{align}
\label{param}
\bm{\xi} &\simeq
 \begin{pmatrix}
  23.54 \sigma \\
  0.321 \pi \\
  0.321 \pi \\
  0
 \end{pmatrix}, \\
\bm{\Xi} &\simeq
 \begin{pmatrix} 
  4.8\thinspace10^{-3} /\sigma^{2}& 0.19 /\sigma\pi & 0.19 /\sigma\pi& -1.8\thinspace10^{-3}/ \sigma\pi \\
  0.19 /\sigma\pi & 1.21 /\pi^{2} & 1.54 /\pi^{2} & 1.25 /\pi^{2} \\
  0.19 /\sigma\pi & 1.54 /\pi^{2} & 1.21 /\pi^{2} & 1.25 /\pi^{2} \\
  -1.8\thinspace10^{-3}/\sigma\pi & 1.25 /\pi^{2} & 1.25 /\pi^{2} & -0.32 /\pi^{2}
 \end{pmatrix},
\label{params}
\end{align}
resulting in the correlation matrix
\begin{align}
\bm{\varrho}^{app}\simeq
 \begin{pmatrix}
  1 & -0.306 & -0.306 & 0.045 \\
  -0.306 & 1 & 0.027 & -0.071 \\
  -0.306 & 0.027 & 1 & -0.071 \\
  0.045 & -0.071 & -0.071 & 1
 \end{pmatrix}.
\label{correff}
\end{align}
It would be unrealistic to try to exactly reproduce all properties of $p_c(\gam)$ through
 an analytical approximation.
Nevertheless, using the resulting expressions from (\ref{papp})-(\ref{params}),
 we can take account of the strong anticorrelation $\varrho_{r \theta_{1,2}}$ between
 $r$ and $\theta_{1,2}$, which is the strongest and most important one in the system;
 compare (\ref{corrc}) and (\ref{correff}).
For the other elements $\varrho_{r\phi}$, $\varrho_{\theta_{1,2} \phi}$, and
 $\varrho_{\theta_{1} \theta_{2}}$ we then obtain stronger deviations.
However, those correlations are smaller than $\varrho_{r\theta_{1,2}}$ \am{at least by a factor $4$} and therefore carry a smaller amount of information about the overall system behavior.

As a total result and in analogy to (\ref{ec}), we obtain from (\ref{papp}) the optimized
 analytical expression $V_{app}$ to model the effective interaction between the mesoscopic particles:
\begin{align*}\label{anal_pot}
V&_{app}(\gam) =   V_{WCA}(r/2a) + V_{FENE}(r) + V_D(\theta_1)
\addtocounter{equation}{1}\tag{\theequation} \\
 &+ V_D(\theta_2) +V_\tau(\phi) +k_BT\sum_{\alpha\beta}
 \Xi_{\alpha\beta}(\alpha-\xi_\alpha)(\beta-\xi_\beta).
\end{align*}
The corresponding expressions and values for the fitting parameters are given by (\ref{WCA}),
 (\ref{fene})-(\ref{etau}), (\ref{param}), and (\ref{params}) as well as  figures \ref{fitfene}-\ref{fittp}. 
Thus, the effective \am{pair} potential is divided into two parts: one-variable and two-variable potentials.
The former are the analytical single-variable \am{pair} potentials derived in section \ref{margpot},
 see (\ref{WCA}) and (\ref{fene})-(\ref{etau}), together with the diagonal $\alpha=\beta$
 terms in the double summation of (\ref{anal_pot}).
The latter are the off-diagonal $\alpha\neq\beta$ terms and take account, to lowest order,
 of the correlations between different mesoscopic degrees of freedom.
Correlations between $r$ and $\theta_{1,2}$ are the dominating ones, leading to such physical
 effects as the wrapping effect introduced in section \ref{wrap}.

\section{Impact of magnetic dipole moments}\label{magneff}
\begin{figure}[]
\centering
  \includegraphics[width=8.6cm]{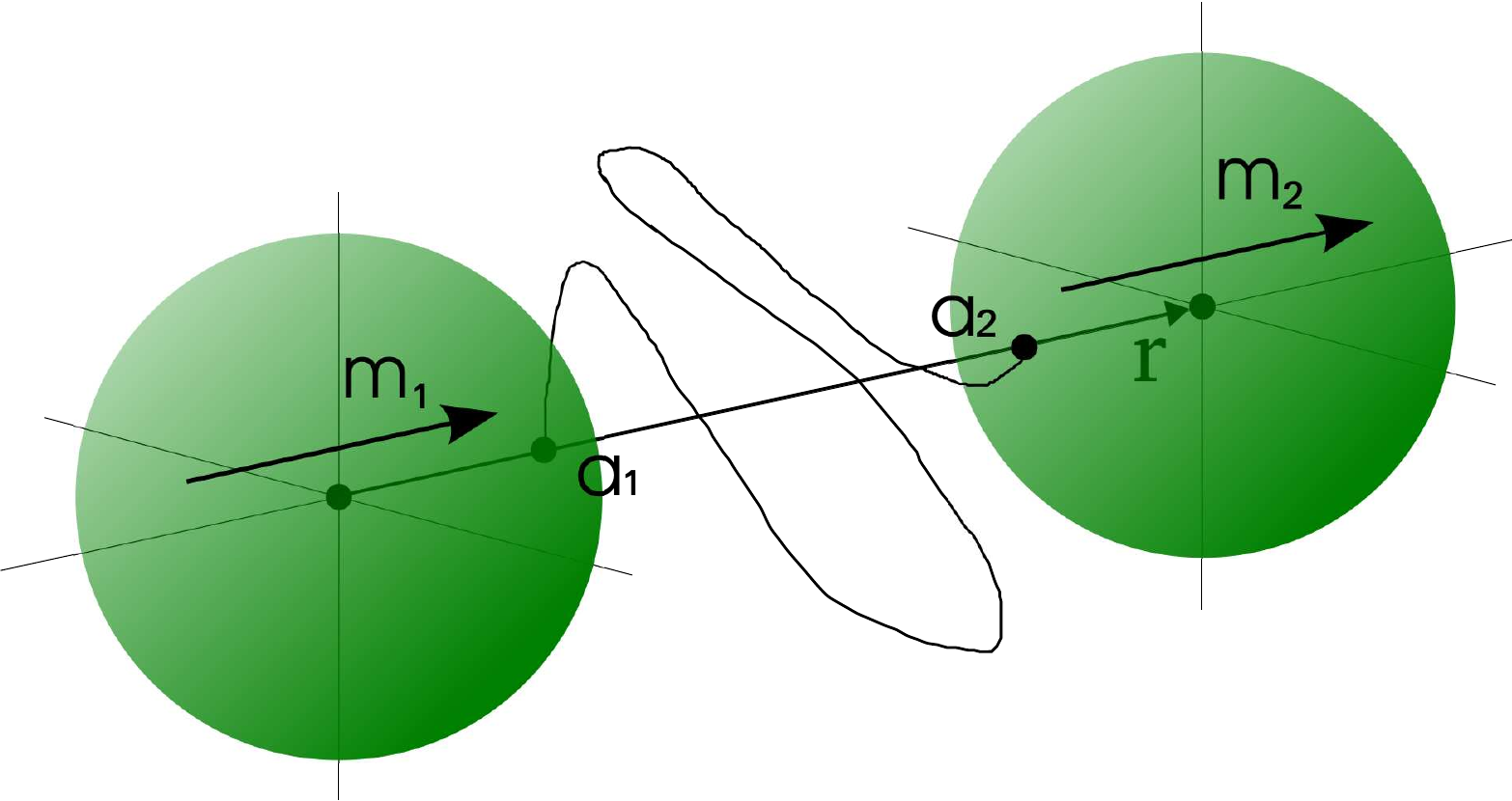}
  \caption{A simplified sketch of the mesoscopic configuration in which the magnetic moments
 are assigned to the particles.
 The depicted orientations $\bm{a}_1$ and $\bm{a}_2$, which identify the anchoring points
 of the polymer chain, correspond to the $\theta_1=\theta_2=0$ and $\phi=0$ configuration.
  The distance between the centers of the mesoscopic particles here corresponds to $r= 26 \sigma$.
 In this configuration, the magnetic moments $\bm{m}_1$ and $\bm{m}_2$ are introduced
 parallel to each other and pointing along the connecting vector $\bm{r}$.}
  \label{mom_init}
\end{figure}

We will now consider how magnetic interactions between the mesoscopic particles modify the physics of the system, in particular the probability densities and other mesoscopic quantities.
For this purpose, we assign to each particle a permanent magnetic dipole moment $\bm{m}_i$ ($i=1,2$),
 in the present work not going into the details of how these could be generated.
We introduce these magnetic moments in the state of highest probability density.
At vanishing magnetic moments, this is the state of minimal effective energy of the mesoscopic system, i.e., following (\ref{ec}), the one that has maximum $p_c(\gam)$ over all the configurations $\gam$.
The maximum occurs for $(r^{M},\theta_1^{M},\theta_2^{M},\phi^{M})\simeq (26\sigma,0.0,0.0,0.0)$,
 and is displayed in figure \ref{mom_init}.
When the mesoscopic particles are in this configuration, we assign to them the two magnetic moments $\bm{m}_1$ and $\bm{m}_2$ that are parallel to each other and point along the connecting vector $\bm{r}$, i.e.~in their orientation of minimal magnetic energy, as depicted in figure \ref{mom_init}.
To first order, this configuration should leave the angular orientations unchanged when increasing the magnetic interaction, but see also the discussion in section \ref{wrap}.
This is just one of the possible orientations that the moments could assume.
However, such a configuration could be achieved with a certain probability, for instance, when the sample is cross-linked in the presence of an external magnetic field that aligns
 the magnetic moments \cite{gunther2012xray,collin2003frozen,varga2003smart,borbath2012xmuct}.
The dipole moments $\bm{m}_1$ and $\bm{m}_2$ are assumed to have equal magnitude
 $m=|\bm{m}_1|=|\bm{m}_2|$ and are rigidly fixed with respect to each particle frame.
We measure the magnetic moments in multiples of $m_0=\sqrt{4\pi k_BT{\left(2a\right)}^3/\mu_0}$,
 where $\mu_0$ is the vacuum magnetic permeability.
Then, for each $\gam$, we can calculate the magnetic dipole interaction energy between the two moments
\begin{align*}
\label{dipdippot}
&V_m(\gam) = \frac{\mu_0}{4 \pi}  \frac{\bm{m}_1\cdot\bm{m}_2 r^2
 -3(\bm{m}_1\cdot\bm{r})(\bm{m}_2\cdot\bm{r})}{r^5} \addtocounter{equation}{1}\tag{\theequation}\\
&= \frac{\mu_0 {m}^2}{4 \pi} \frac{-2 \cos(\theta_1^m) \cos(\theta_2^m)
 +\sin(\theta_1^m)\sin(\theta_2^m)\cos(\phi^m)}{r^3}. 
\end{align*}
$\theta_i^m$ indicates the angle between $\bm{m}_i$ and $\bm{r}$, while $\phi^m$ is the angle
 between the projections of $\bm{m}_1$ and $\bm{m}_2$ on a plane perpendicular to $\bm{r}$.
These quantities can be expressed in the $\gam$ variables once the orientations
 of $\bm{m}_1$ and $\bm{m}_2$ with respect to the particle frames are set by the protocol
 we described above.

\am{For non-magnetic polymer chains, the potential acting on the mesoscopic level $V(\gam)$ is separable into a sum of magnetic and non-magnetic interactions $V(\gam) = V_c(\gam) +V_m(\gam)$.
Consequently, magnetic effects do not modify the contributions of the polymer chain to $V_c(\gam)$, as is further explained in \am{appendix~\ref{app1}}.}

This corresponds to a factorization of the probability densities $p_c(\gam)$ and $p_m(\gam)$,
 where the latter is defined via the Boltzmann factor
\begin{equation}\label{pm}
 p_m(\gam) = \exp\left[ -\beta V_m(\gam) \right]
\end{equation}
in analogy to (\ref{ec}).
Therefore the total probability density becomes
\begin{equation}\label{ptot}
 p(\gam)= \frac{e^{-\beta [V_c(\gam)+V_m(\gam)]}}{Z(m)} = \frac{p_c(\gam) p_m(\gam)}{Z(m)},
\end{equation}
 where $Z(m)=\int p_c(\gam) p_m(\gam) \d\gam$ is the partition function describing the system
 for non-vanishing magnetic moments.
We can calculate averages on the system with magnetic interactions by
\begin{equation}\label{avgvar}
 \langle \cdot \rangle  = \int \cdot \ \ p(\gam)\ \d\gam
= \frac{\int \cdot \ \ p_c(\gam)\ p_m(\gam)\ \d\gam}{Z(m)}.
\end{equation}

\begin{figure}[]
\centering
  \includegraphics[width=8.6cm]{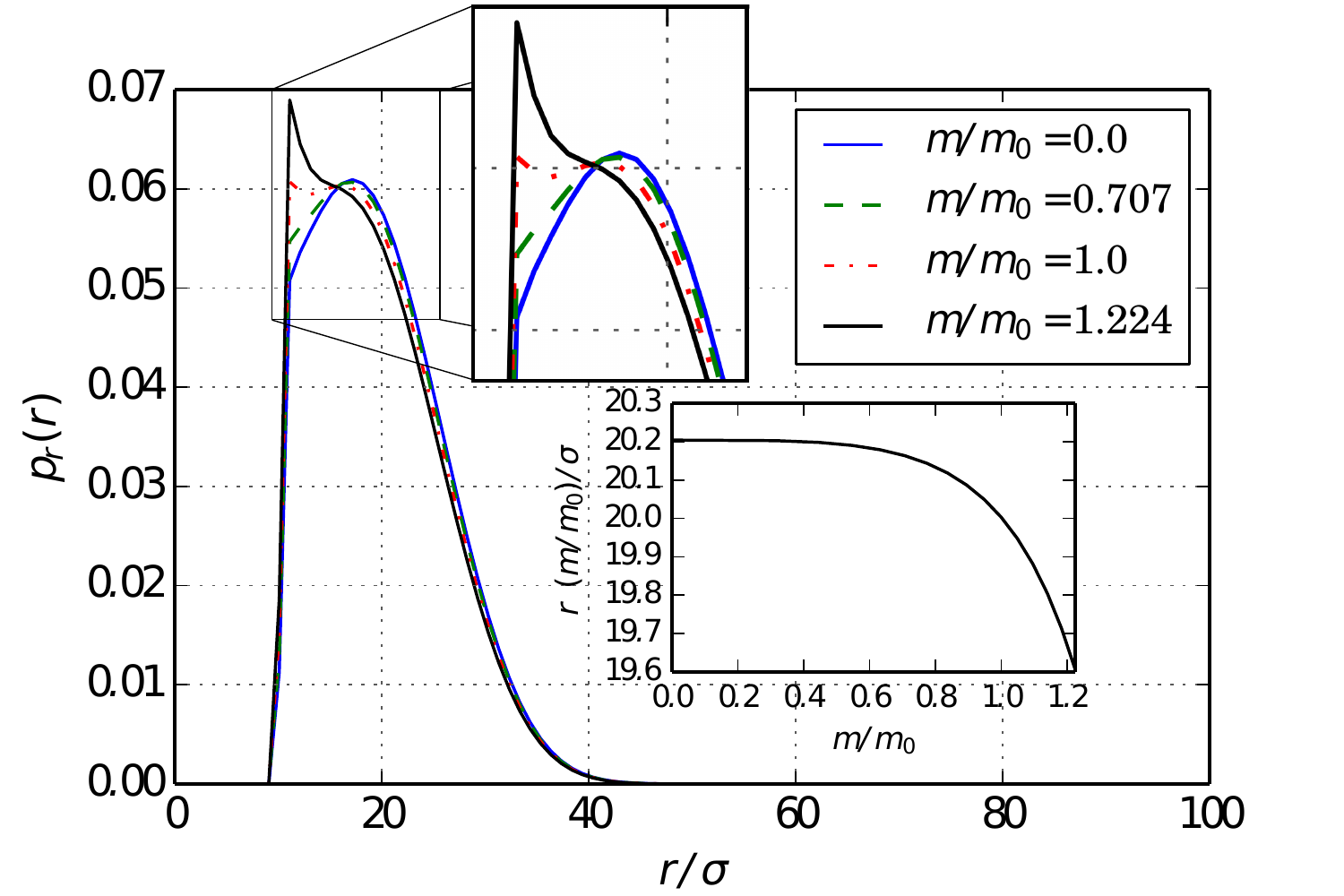}
  \caption{Nonvanishing permanent magnetic dipole moments of the mesoscopic particles and their impact on the system.
We plot the marginal probability density $p_r$ for the distance $r$ between the particles
 for different values of the reduced magnetic moment $m/m_0$.
Inset: average particle distance $\langle r \rangle$ as a function of the reduced magnetic moment $m/m_0$.}
  \label{prm}
\end{figure}

The single-variable marginal probability density $p_\alpha(\alpha)$, $\alpha=r,\theta_1,\theta_2,\phi$,
 is again defined as $p(\gam)$ integrated over all the $\gam$ variables except for $\alpha$.
This is the same procedure as described in section \ref{stat} but substituting $p_c(\gam)$ with $p(\gam)$. 
As can be seen from figure \ref{prm}, ${p}_r(r)$ shows an increase of the probability to find the particles close together when $m$ is increased.
A peak builds up at small $r$ because the magnetic energy tends to $V_m(r\rightarrow 0)\rightarrow -\infty$.
Although the magnetic spheres attract each other, a collapse is prevented by the WCA-potential,
 which becomes effective at $r \lesssim 11\sigma$ and for $r\rightarrow 0$ diverges faster
 to $+\infty$ than the magnetic energy to $-\infty$.
For $m=m_0$, the presence of a double peak in ${p}_r(r)$ could be connected to a hardening transition of the kind
 described in \cite{annunziata2013hardening}.
Due to the mutual magnetic attraction between the parallel dipoles we expect the average
 distance $\langle r \rangle $ to decrease with increasing $m$, and indeed it does so,
 as can be seen from the inset in figure \ref{prm}.

\begin{figure}[]
\centering
  \includegraphics[width=8.6cm]{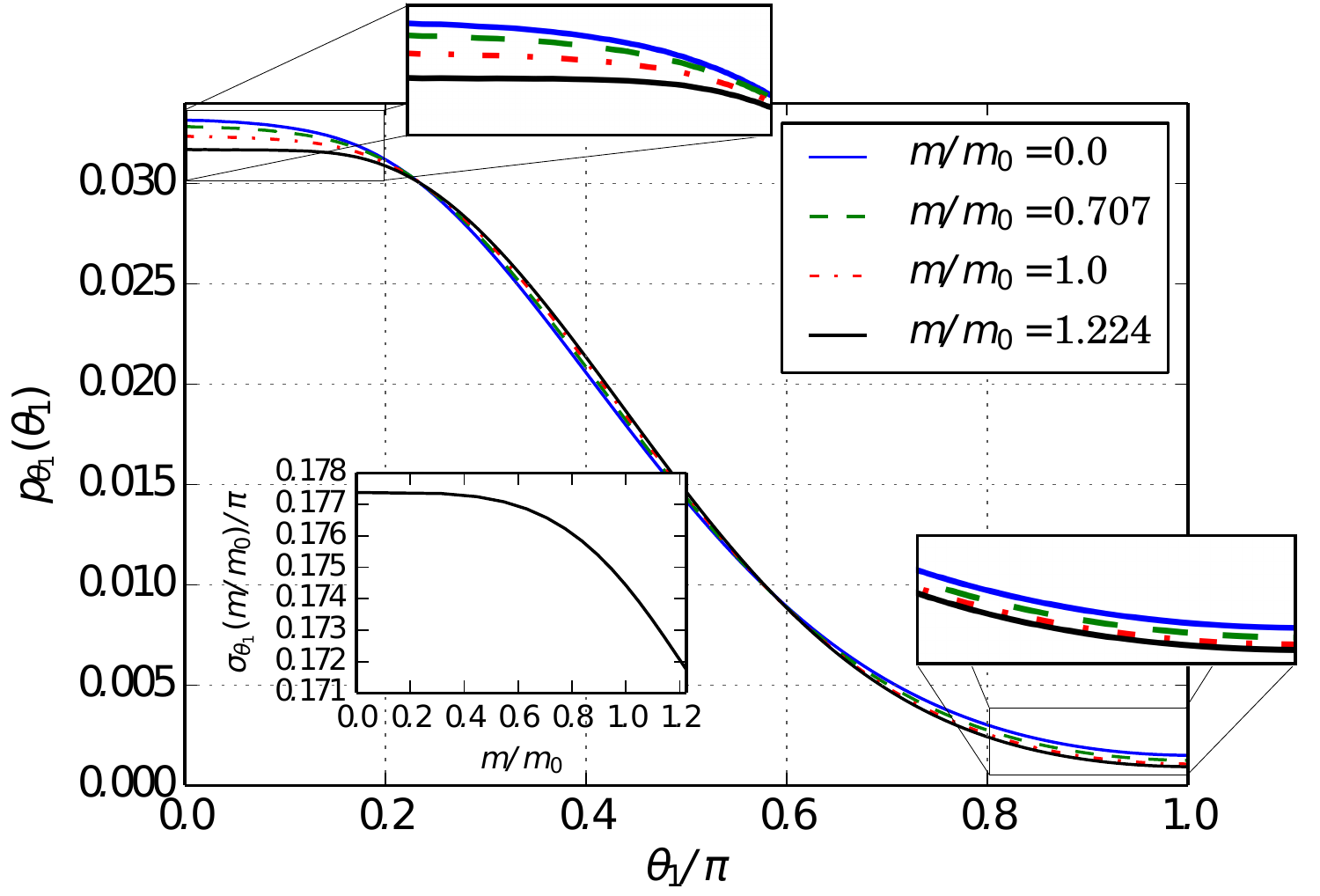}
  \caption{Marginal probability density $p_{\theta_1}$ for the angle $\theta_1$
 (very similar for $\theta_2$) for increasing reduced magnetic moment $m/m_0$.
Inset: standard deviation of $\theta_1$, $\sigma_{\theta_1} =
 \sqrt{\langle{ {\left(\theta_1 -\langle\theta_1\rangle \right)}^2 }\rangle}$
 (very similar for $\theta_2$) as a function of $m/m_0$.}
  \label{prth}
\end{figure}

The changes in the angular distributions for $\theta_1$ and $\theta_2$ due to the
 magnetic interaction are illustrated in figure \ref{prth}.
For the two angles $\theta_1$ and $\theta_2$ the distributions $p_{\theta_1}(\theta_1)$ and
 $p_{\theta_2}(\theta_2)$ are similar, and the behavior for varying $m$ is approximately the same.
Therefore, we only display the results for $p_{\theta_1}(\theta_1)$.
The magnetic moments tend to align in parallel along the connecting axis $\bm{r}$, corresponding to their absolute energy minimum.
Since we introduced the magnetic moments in that configuration for $\theta_1=\theta_2=0$, for any fixed $\theta_1$
 the minimum of $V_m$, and therefore the maximum of $p_m$, is located at $\theta_2 = \theta_1$ and vice versa.
As a consequence, the $\theta_{1}=\theta_2=0$ configuration leads to the same magnetic interaction
 energy as $\theta_1=\theta_2=\pi$.
Concerning only magnetic interactions, both configurations show the same probability $p_m$.
Integrating out $r$, $\theta_2$, and $\phi$ from $p_m(\gam)$, the shape of the resulting magnetic probability density
 for $\theta_1$ is symmetric around  $\theta_1=\pi/2$ with one peak at $\theta_1=0$ and one at $\theta_1=\pi$.
Therefore, coupling magnetic ($p_m$) and non-magnetic ($p_c$) contributions, we find that
 some probability shifts to higher values of $\theta_1$ due to the magnetic interactions.
However, we find the standard deviation of $p_{\theta_1}(\theta_1)$ to decrease with increasing $m$
 (see the inset of figure \ref{prth}), meaning that the particles become less likely to rotate along
 the $\theta_1$ (or likewise $\theta_2$) direction.

\begin{figure}[]
\centering
  \includegraphics[width=8.6cm]{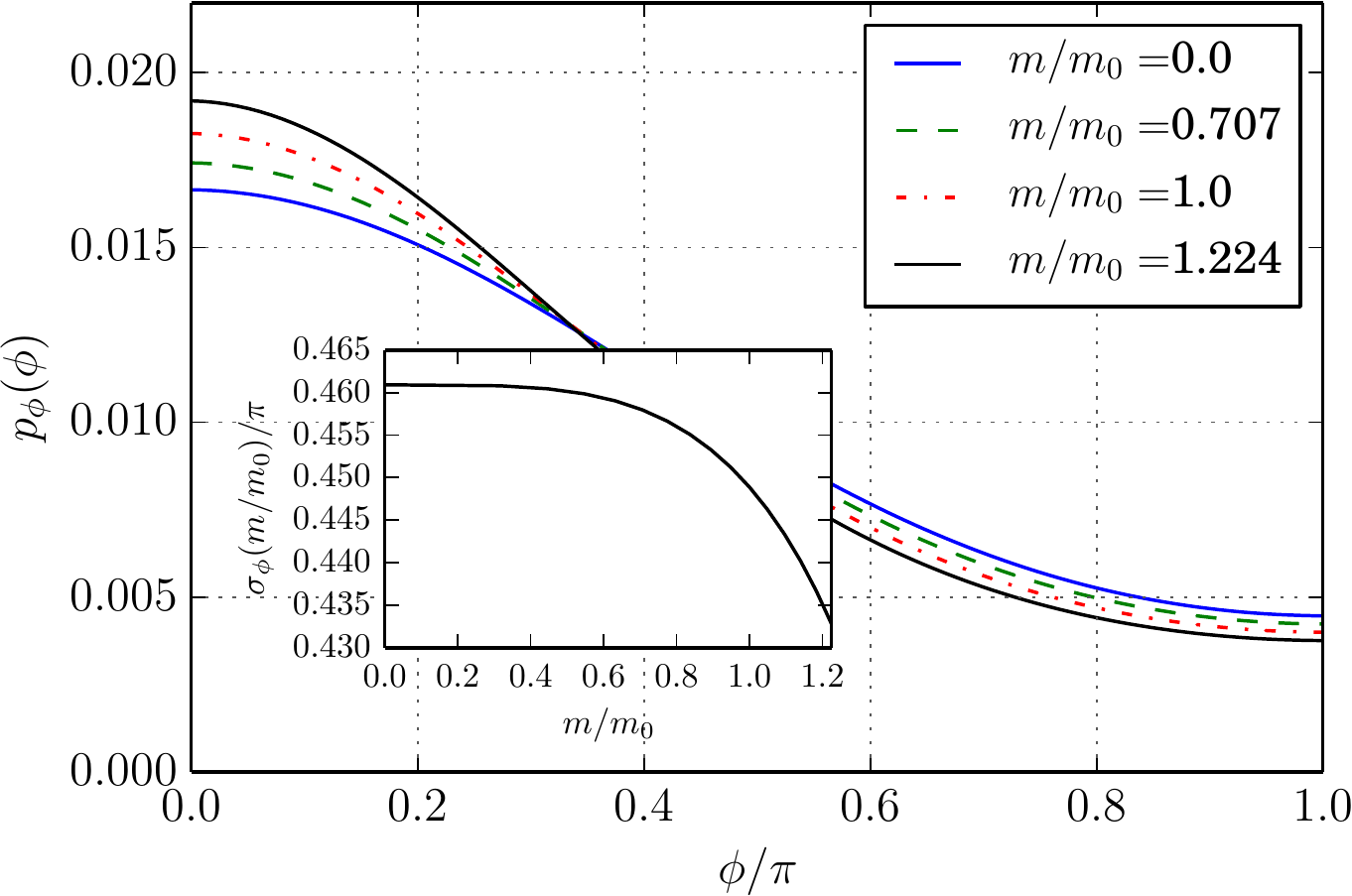}
  \caption{Marginal probability density $p_\phi$ for the angle $\phi$ describing the relative torsion
 between the particles, for increasing reduced magnetic moment $m/m_0$.
 Inset: standard deviation of $\phi$, $\sigma_{\phi} = \sqrt{\langle{ {\left(\phi -\langle\phi\rangle \right)}^2 }\rangle}$
 as a function of $m/m_0$.}
  \label{prphi}
\end{figure}

The same calculation for $\ptil_\phi(\phi)$ shows that the particles also become less likely
 to rotate around the connecting vector $\bm{r}$.
The probability density of $\phi\simeq 0$ rises and sharpens, as we can see in figure \ref{prphi},
and the standard deviation (see the inset of figure \ref{prphi}) decreases, confirming quantitatively
 the sharpening of $\ptil_\phi(\phi)$.

\section{Thermodynamic Properties}\label{thermod}
Finally, we provide the connection to the thermodynamics of our canonical system \am{and demonstrate the influence of the magnetic interactions}.
With the partition function $Z(m)$ as in (\ref{ptot}), we can calculate the overall free energy as
 $F(m)=-k_BT\ln[Z(m)]$, the internal energy of the system as
\begin{equation}\label{inten}
U(m) = \langle V \rangle(m) = \frac{\int V(\gam) p_c(\gam) p_m(\gam) \d\gam}{Z(m)}
\end{equation}
with $V(\gam)$ defined before (\ref{pm}), and the entropy as $S(m) =[ U (m)-F(m)]/T$.

\begin{figure}[]
\centering
  \includegraphics[width=8.6cm]{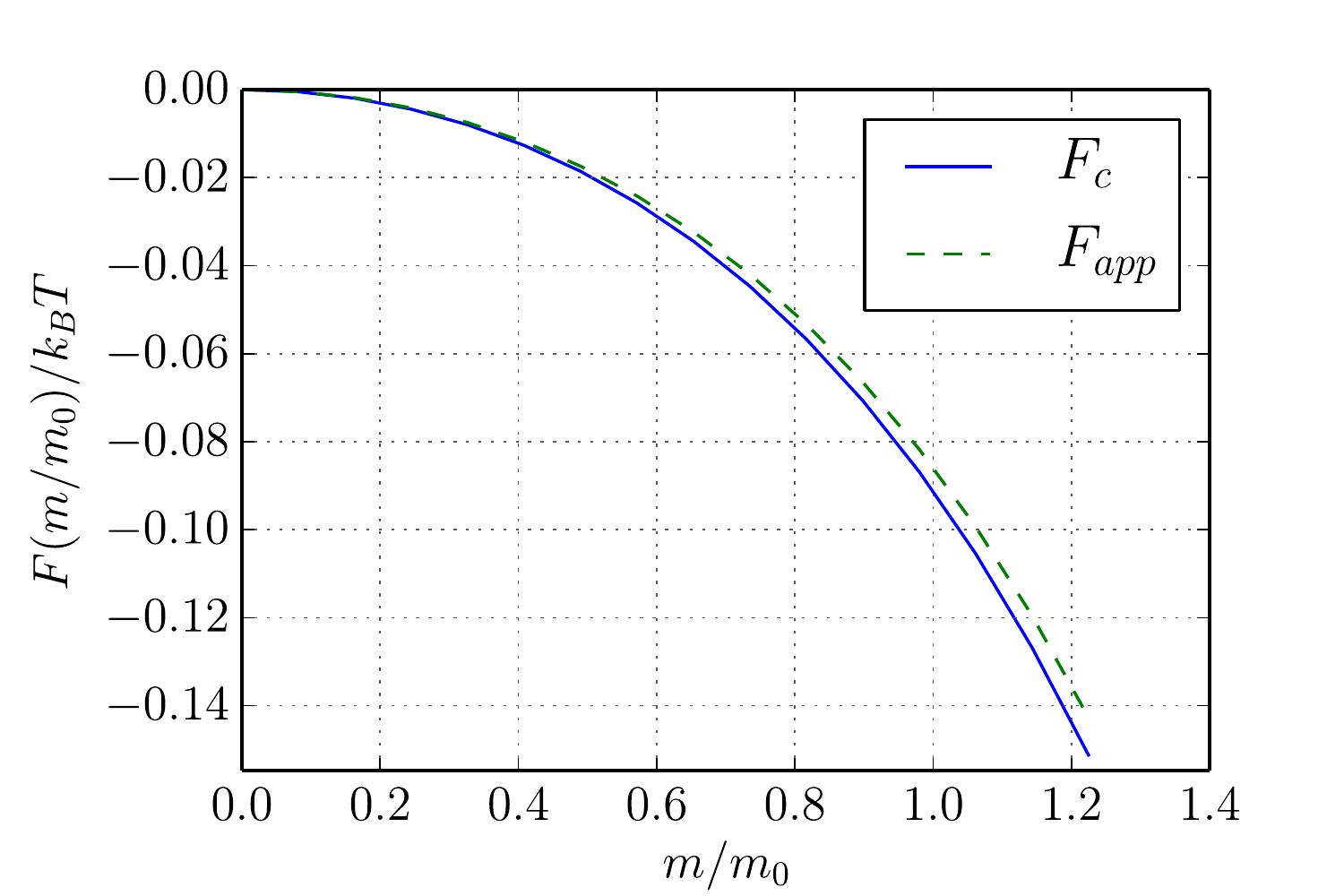}
  \caption{Free energy $F(m)=-k_BT\ln[Z(m)]$ as a function of the reduced magnetic moment $m/m_0$.
The partition function $Z(m)$ is calculated once using $p_c(\gam)$ and once using $\papp(\gam)$.
The result from the microscopic simulation is labeled with ``$F_c$'' and the one from the analytical
 approximation with ``$F_{app}$''.
Note that for $m=0$ the free energies are equal.}
  \label{freeen}
\end{figure}

\begin{figure}[]
\centering
  \includegraphics[width=8.6cm]{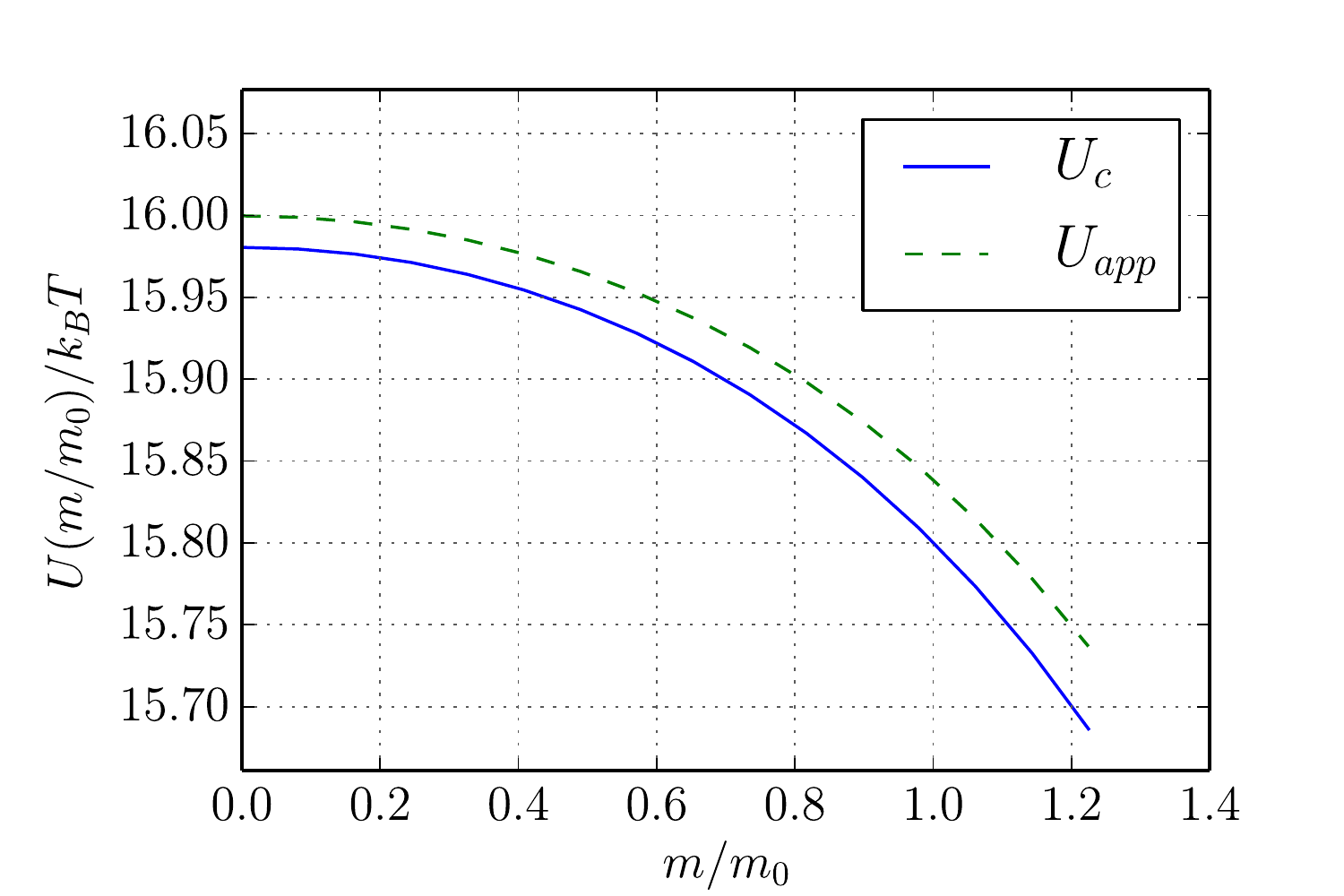}
  \caption{Internal energy $ U (m)$ as a function of the reduced magnetic moment
 $m/m_0$, calculated according to (\ref{inten}).
The curve labeled with ``${ U }_{c}$'' shows the result from the microscopic data
 using (\ref{inten}), whereas in the one labeled with ``${ U }_{app}$'' $p_c(\gam)$
 has been replaced by $\papp(\gam)$.}
  \label{meanen}
\end{figure}

\begin{figure}[]
\centering
  \includegraphics[width=8.6cm]{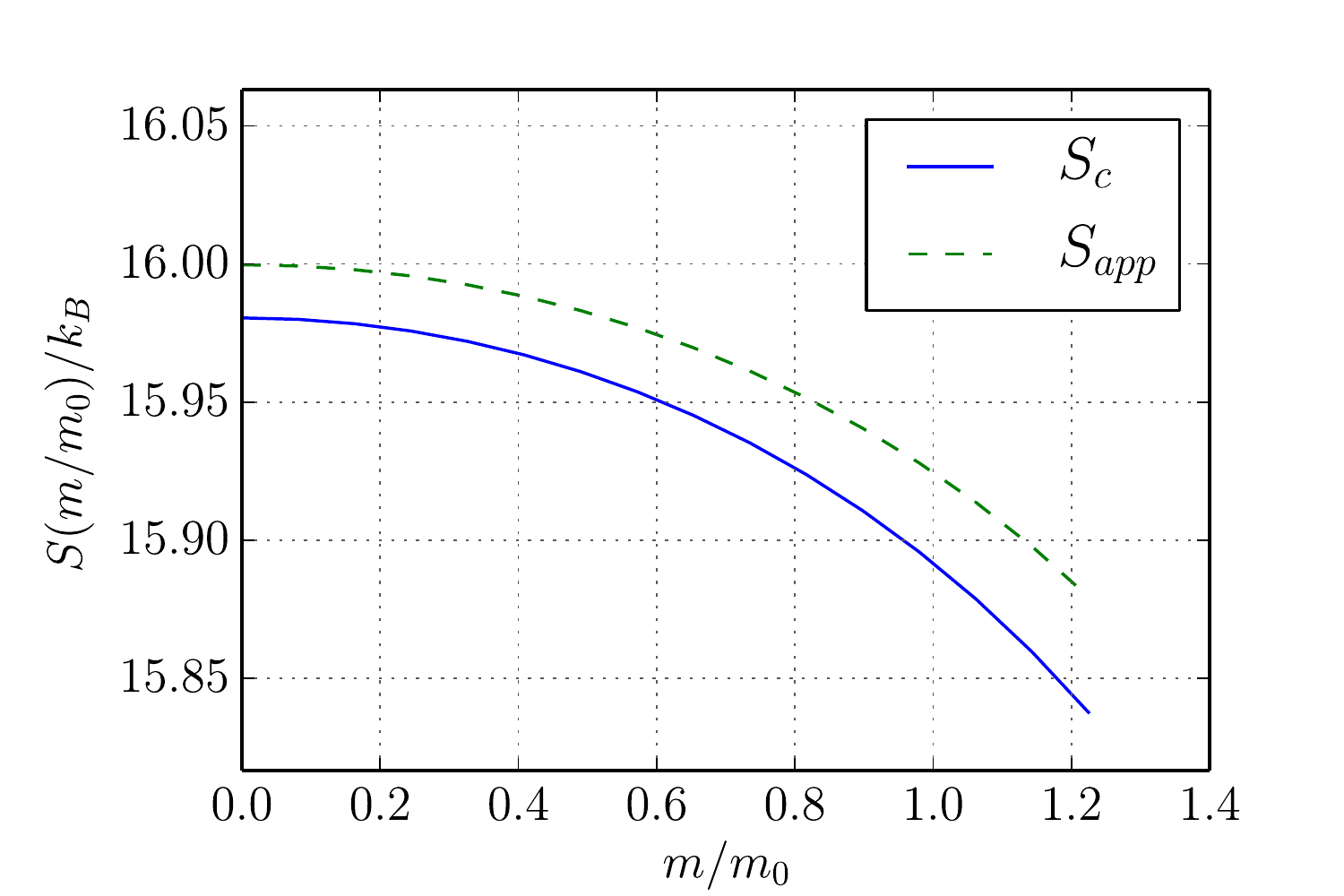}
  \caption{Entropy $S (m)=[U (m)-F(m)]/T$ as a function of
 the reduced magnetic moment $m/m_0$ calculated once using $p_c(\gam)$ from the microscopic simulation
 (labeled with ``${S}_{c}$''), and once replacing $p_c(\gam)$ by $\papp(\gam)$
 (labeled with ``${S}_{app}$''). }
  \label{entropy}
\end{figure}

To test the validity of our effective potential scheme using the coupling approximation
 described in section \ref{effpot}, we calculate the thermodynamic quantities using both probability
 densities $p_c(\gam)$ and $\papp(\gam)$ 
and compare the results.
At vanishing magnetic moment, the free energy of the system is the same for both probability densities.
This is expected, because it is a direct consequence of the normalization condition:
 $Z(m=0)=Z_c=\int p_c(\gam) \d\gam=\int \papp(\gam) \d\gam=1$ by construction of $\papp(\gam)$.
For non-vanishing magnetic moments, the partition functions obtained from the two probability densities (and thus the corresponding free energies) start to deviate from each other 
 because the integral $\int p_c(\gam) p_m(\gam) \d\gam$ is, in general, different from
 $\int \papp(\gam) p_m(\gam) \d\gam$.
With increasing magnetic interaction, the difference between $p_c(\gam)$ and $\papp(\gam)$
 becomes more important and, as shown in figure \ref{freeen},
 leads to a slightly increasing deviation in the free energies $F(m)$ calculated in both ways:
 at $m\simeq 1.224 m_0$ they already differ by $\sim 5.6\%$.
Analogously, we can compare the internal energies of the system, shown in figure \ref{meanen}:
 at $m=0$ the error due to the probability density approximation is $\sim 0.12\%$ of the exact value,
 rising up to $\sim 0.32\%$ at $m\simeq 1.224 m_0$.
A similar deviation follows for the entropy, see figure \ref{entropy}, where, however, the error
 at $m\simeq1.224m_0$ increases to $\sim 0.26\%$ of the exact value.
Overall, however, the relative errors remain small, which confirms the validity and viability
 of our coarse-grained effective model potential in (\ref{anal_pot}).

To summarize the effect of the magnetic dipoles on the system, we may conclude that the particles
 are driven towards each other.
In other words, the average distance between them decreases (see figure \ref{prm}) and
 the distributions for the particle separation and for the angular degrees of freedom sharpen
 (see figures \ref{prm}, \ref{prth}, and \ref{prphi}).
This reflects a decrease in entropy, which becomes possible due to the gain in magnetic interaction energy.
Indeed, the calculated entropy decreases with increasing magnetic interactions (see figure \ref{entropy}), which is achieved by the decreasing free energy and internal energy (see figures \ref{freeen} and \ref{meanen}, respectively).

\section{Conclusions}\label{conclusions}
Most of the previous studies on ferrogels describe the polymer matrix as a continuous material
 \cite{szabo1998shape,zubarev2012theory,ivaneyko2012effects,wood2011modeling}
 or represent it by springs connecting the particles \cite{annunziata2013hardening,
dudek2007molecular,sanchez2013effects,tarama2014tunable,pessot2014structural,cerda2013phase},
 but only a few resolve explicitly the polymeric chains \cite{weeber2012deformation,weeber2015ferrogels}.
In particular, a link between such microscopic chain-resolved calculations and the expressions for the investigated
 mesoscopic model energies has so far been missing.
We have outlined in the present work a way to connect detailed microscopic simulations to a coarse-grained
 mesoscopic model.

This manifests a step into the direction of bridging the scales in material modeling.
Starting from  microscopic simulations considering explicitly an individual polymer chain
 connecting two mesoscopic particles, we specified effective mesoscopic \am{pair} potentials 
 by fitting analytical model expressions.
In this way, we were able to optimize a coarse-grained mesoscopic model description
 on the basis of the input from the explicit microscopic simulation details.
Furthermore, we have shown that correlations between the mesoscopic degrees of freedom must be taken into
 account to provide a complete picture of the physics of the system.
Moreover, we have examined the effect of \am{a} magnetic interaction, finding \am{a tightening of} the system
 by reducing the average distance between the magnetic particles and \am{by reducing} the rotational fluctuations
 around the configuration of highest probability density.

\am{
Our system consisted of only two mesoscopic particles, for which we derived the corresponding effective pair interaction potential.
However, using this pair potential in a first approximation, two- and three-dimensional structures can be built up, similarly to elastic network structures generated using pairwise harmonic spring interactions between the particles \cite{pessot2014structural,tarama2014tunable}.
Including as a first approach magnetic interactions between neighboring particles only, the different angles between the magnetic moment of a particle and the anchoring points of the polymer chain to its different neighbors must be taken into account.
Yet, using the reduced picture of pairwise mesoscopic interactions, it should be possible to reproduce for example previously observed deformational behavior in external magnetic fields for two- and three-dimensional systems \cite{weeber2012deformation,weeber2015ferrogels}. 
}

The scope of our approach is a first attempt of scale-bridging in modeling ferrogels and magnetic elastomers.
Naturally, the procedure can be improved in many different ways.
For example, we would like to study a system with multiple chains connecting the magnetic particles, with
 anchoring points randomly distributed over the surfaces of the particles.
Such a development would eliminate artificial symmetries in the model and be another step towards
 the description of real systems.
Furthermore, to study more realistic systems, interactions between more than two mesoscopic particles
 would need to be considered.
Likewise, also the effect of interactions between next-nearest neighbors connected
 by polymer chains could be included. 
\am{The last two points imply a step beyond the reduced picture of considering only effective pairwise interactions between the mesoscopic particles.} 
Finally, via subsequent procedures of scale-bridging from the mesoscopic to the macroscopic
 level \cite{menzel2014bridging}, a connection between microscopic details and macroscopic
 material behavior may become attainable for magnetic gels.

\begin{acknowledgments}
The authors thank the Deutsche Forschungsgemeinschaft for support of this work through
 the priority program SPP 1681.
RW and CH acknowledge further funding through the cluster of
excellence EXC 310, SimTech, and are grateful for the access to the computer facilities of
the HLRS and BW-Unicluster in Stuttgart.
\end{acknowledgments}

\appendix

\section{Dependence on microscopic system parameters}\label{app2}
\am{In sections \ref{margpot} and \ref{effpot} we derived an analytical approximation for the effective mesoscopic model potential describing our system.
Here, we test how the mesoscopic model parameters depend on the microscopic system parameters. In particular, we performed additional microscopic simulations with varied radius $a$ of the mesoscopic particles and varied number $N$ of beads forming the connecting polymer chain.
}
\begin{figure}
\centering
  \includegraphics[width=8.6cm]{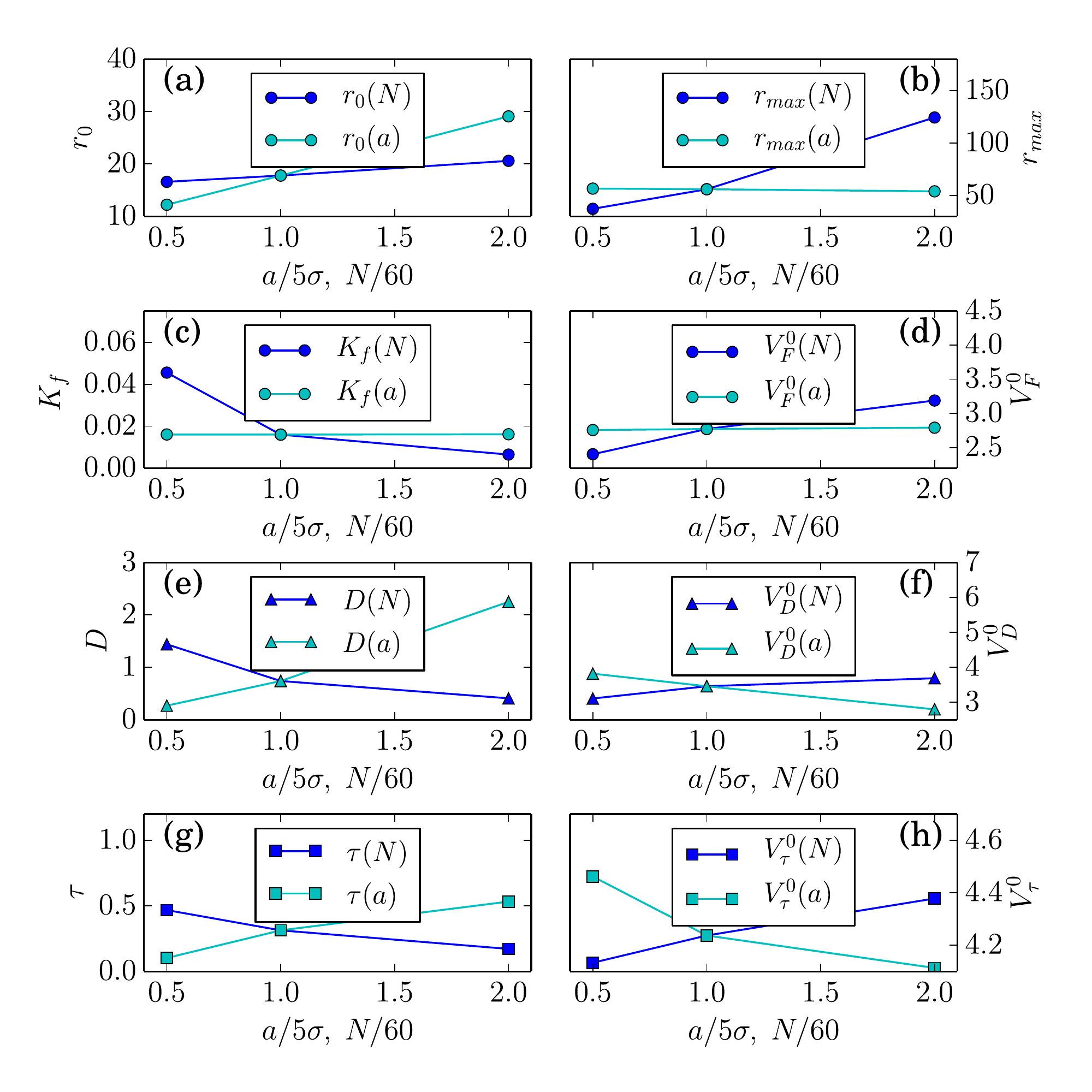}
  \caption{\am{Effect of varying microscopic system parameters on the resulting mesoscopic model parameters. In the microscopic simulations, we varied the mesoscopic particle radius $a$ and the number of beads $N$ forming the connecting polymer chain. Fits of the single-variable mesoscopic model potentials (\ref{fene})--(\ref{etau}) to the microscopic simulation data lead to the presented mesoscopic parameter values. (a--d, $\Circle$) Model parameters for the FENE potential (\ref{fene}); 
(e,f, $\triangle$) model parameters for the angular $\theta_1$- and $\theta_2$-potential (\ref{ed}); (g,h, $\square$) model parameters for the torsional $\phi$-potential (\ref{etau}).
Values for $\theta_0$ and $\phi_0$ vanished in all cases and are not shown.  
The data points for $a/5\sigma=1$ and $N/60=1$ correspond to the results presented in figures \ref{fitfene}--\ref{fittp}.}}
  \label{param_scal}
\end{figure}
\begin{figure}
\centering
  \includegraphics[width=8.6cm]{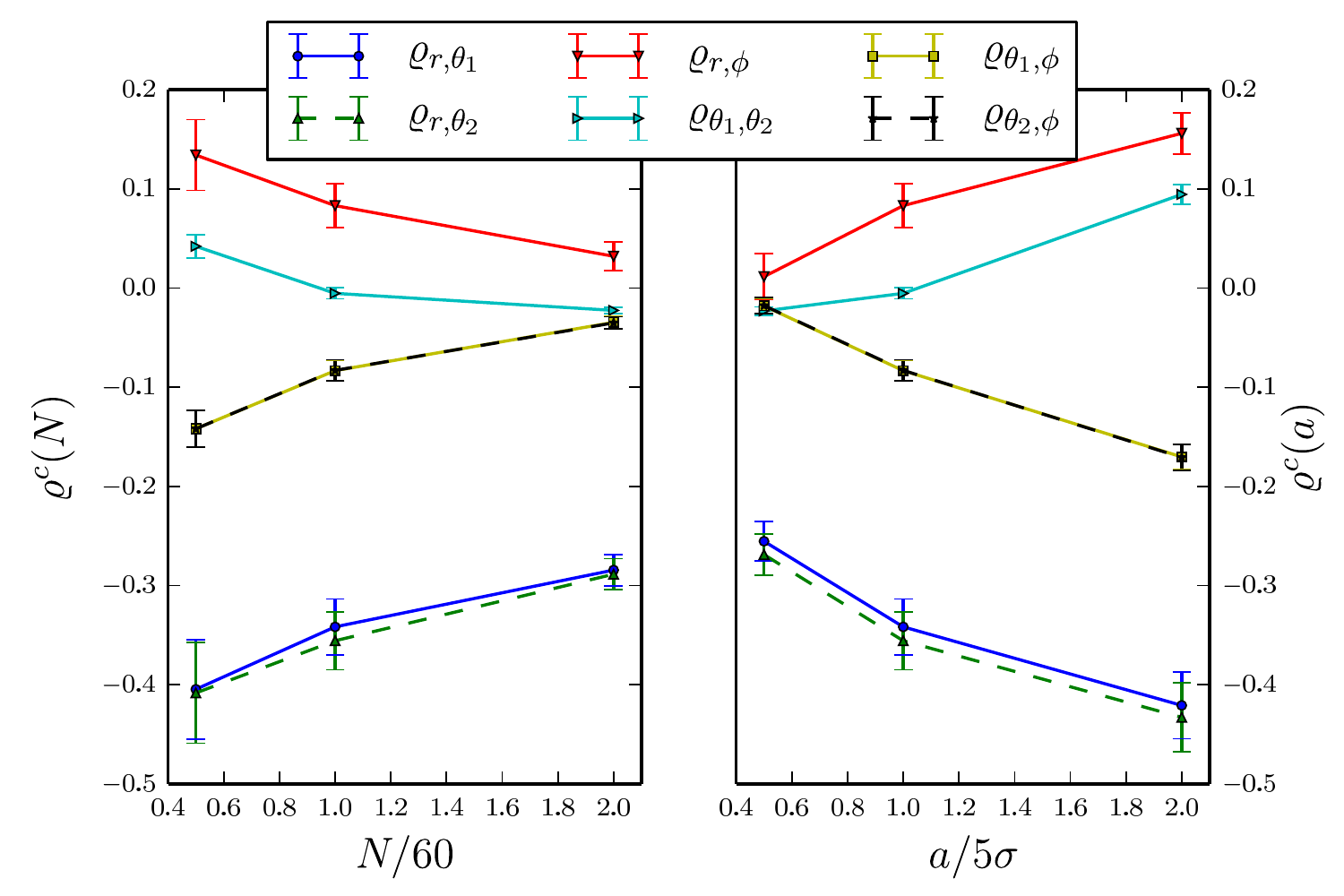}
  \caption{\am{Elements of the correlation matrix $\bm{\varrho}$ for different values of $a$ and $N$.
The data points for $a/5\sigma=1$ and $N/60=~1$ correspond to the results \am{presented in (\ref{corrc})}.} }
  \label{corr_scal}
\end{figure}

\am{In each case, we repeated the analysis of sections \ref{stat}, \ref{margpot}, and \ref{effpot} by fitting the mesoscopic single-variable model potentials (\ref{fene})--(\ref{etau}) to the microscopic simulation data.
This reveals the trends in the dependences of the coefficients in the mesoscopic model potentials on the parameters $a$ and $N$, as depicted in figure~\ref{param_scal}.}
\am{From the trend of the resulting model parameters $K_f$, $D$, and $\tau$ in figure~\ref{param_scal} (c,e,g) we conclude that the potentials tend to become stiffer as the chain becomes shorter or -- at least for the rotational degrees of freedom -- the mesoscopic particles become larger (see \cite{suppl} for corresponding fitting curves).
}

\am{Moreover, we have calculated the trend in the correlations between the mesoscopic degrees of freedom for varying values of $a$ and $N$, see figure \ref{corr_scal}.
We find that the magnitude of the correlations decreases with increasing $N$ or decreasing $a$. Thus, quite intuitively, the coupling between the $\gam$ variables tends to decrease with longer chains or smaller mesoscopic particle sizes.
}

\section{Separability of the Hamiltonian and Consequences for Coarse-Graining}\label{app1}
\am{In section~VIII we have included the magnetic interactions analytically on the mesoscopic level.
They had not been included in the microscopic simulations.
This procedure is possible due to a separability of the magnetic and non-magnetic effects which as a consequence implies a factorization of the corresponding probability.
Thus the contribution of the polymer chain needs to be simulated explicitly.
The contribution of any interaction acting solely on the mesoscopic particles can be exactly taken into account separately afterwards.}

\am{This argument relies on the separability of the Hamiltonian into a sum of mesoscopic and microscopic parts, as well as on the fact that magnetic interactions affect the mesoscopic part only.
We write down the Hamiltonian of the system as
\widetext
\begin{align}
 \mathcal{H}(\gam,\Gam) &= \mathcal{T}_{meso}(\gam) + V_{m}^{\A,\B}(\gam) + V_{WCA}^{\A,\B}(\gam) \label{mesopart} \\
 &+ \mathcal{T}_{micro}(\Gam) + \sum_{n=1}^{N} \left[ V_{WCA}^{\A,n}(\gam,\Gam) +V_{WCA}^{n,\B}(\gam,\Gam) \right] + \sum_{n< n'}^{N} V_{WCA}^{n,n'}(\Gam) \label{microkinwca} \\
 &+ V_{H}^{\A,1}(\gam,\Gam) + V_{H}^{N,\B}(\gam,\Gam) \label{microharm} + \sum_{n=1}^{N-1}V_{H}^{n,(n+1)}(\Gam).
\end{align}
\endwidetext
Similarly to the main text, let us here indicate with $\gam$ the degrees of freedom (now velocities included) of the mesoscopic particles and with $\Gam$ the ones of the microscopic particles that build up the chain.
For brevity we here label the mesoscopic particles by $\A$ and $\B$ and the microscopic ones
 by the discrete indices $n,n'=1,\dots,N$.
Moreover, we denote by $V_{(\ )}^{p,q}(\gam,\Gam)$ the corresponding interaction between particles $p$ and $q$, where the explicit expressions of $V_{WCA}$, $V_H$, and $V_m$ are as described in (\ref{WCA}), (\ref{harmspring}), and (\ref{dipdippot}).
$\mathcal{T}_{meso}(\gam)$ and $\mathcal{T}_{micro}(\Gam)$ indicate, respectively, the kinetic
 energies of the mesoscopic and microscopic particles.
Therefore, $\mathcal{H}(\gam,\Gam)$ is separable into
\begin{equation}
 \mathcal{H}(\gam,\Gam) = \mathcal{H}_{meso}(\gam) +\mathcal{H}_{micro}(\gam,\Gam),
\end{equation}
 where $\mathcal{H}_{meso}(\gam)$ contains the terms in line~(\ref{mesopart})
 and $\mathcal{H}_{micro}(\gam)$ is composed of the terms in lines~(\ref{microkinwca}) and (\ref{microharm}).}

\am{Since we work in the canonical ensemble, the physics of the system derives from the partition function
\begin{align}
 \mathcal{Z}(m)&= \int \e^{-\beta \mathcal{H}(\gam,\Gam)} \, \d\gam \, \d\Gam \\ \nonumber
&=\int \e^{-\beta \mathcal{H}_{meso}(\gam)} \, \e^{-\beta \mathcal{H}_{micro}(\gam,\Gam)} \, \d\gam \, \d\Gam,
\end{align}
with $\beta=1/k_BT$.
Coarse-graining means to integrate out the microscopic degrees of freedom $\Gam$, so we rearrange
\begin{align}
\label{CG}
 &\mathcal{Z}(m) = \int \e^{-\beta \mathcal{H}_{meso}(\gam)} \left[ \int \e^{-\beta \mathcal{H}_{micro}(\gam,\Gam)} \d\Gam \right] \d\gam \\ \nonumber
 &\ \ =\int \e^{-\beta \mathcal{H}_{meso}(\gam)} \mathcal{Z}_{micro}(\gam) \d\gam \\ \nonumber
 &=\int \e^{-\beta V_{m}^{\A,\B}(\gam) } \e^{-\beta \left[\mathcal{T}_{meso}(\gam)+V_{WCA}^{\A,\B}(\gam)\right] } \mathcal{Z}_{micro}(\gam) \, \d\gam.
\end{align}
The connection to the probability density $p_c(\gam)$ is given by
\begin{equation}
 p_c(\gam)= \frac{ \mathcal{Z}_{micro}(\gam)\e^{-\beta \left[\mathcal{T}_{meso}(\gam)+V_{WCA}^{\A,\B}(\gam)\right]}}{\mathcal{Z}(m=0)}.
\end{equation}
}

\am{Since the magnetic interactions only affect the mesoscopic particles, see line~(\ref{mesopart}), it is solely contained in $\mathcal{H}_{meso}(\gam)$.
Therefore, the microscopic Hamiltonian $\mathcal{H}_{micro}(\gam,\Gam)$ for a specific fixed configuration $\gam$ of the mesoscopic particles does not explicitly depend on the magnetic moments.
The magnetic effects do not modify the physics of the polymer chain for a given configuration of the mesoscopic particles.
The underlying physical reason is that the polymer chain does not contain magnetic components.}

\am{As we can see, the magnetic interactions only appear on the mesoscopic level of the final partition function $\mathcal{Z}$.
The magnetic interactions are simply included by multiplying on the mesoscopic level with the additional probability factor $\exp[-\beta V_m^{A,B}(\gam)]$
The remaining part of the integrand that contains the information drawn from the MD simulations, is not affected.}

\am{Explicitly introducing the magnetic moments in the microscopic simulation would therefore not affect the final outcome.
Thus, it is sufficient to determine the effects of the polymer chain through MD simulations and add
 the magnetic interactions analytically in a subsequent step.}


\begin{thebibliography}{10}

\bibitem{strobl1997physics}
G.~Strobl.
\newblock {\em The Physics of Polymers}.
\newblock Springer Berlin / Heidelberg, 2007.

\bibitem{klapp2005dipolar}
S.~H.~L. Klapp.
\newblock Dipolar fluids under external perturbations.
\newblock {\em J. Phys.: Condens. Matter}, 17(15):R525, 2005.

\bibitem{huke2004magnetic}
B.~Huke and M.~L{\"u}cke.
\newblock Magnetic properties of colloidal suspensions of interacting magnetic
  particles.
\newblock {\em Rep. Prog. Phys.}, 67(10):1731, 2004.

\bibitem{rosensweig1985ferrohydrodynamics}
R.~E. Rosensweig.
\newblock {\em Ferrohydrodynamics}.
\newblock Cambridge University Press, Cambridge, 1985.

\bibitem{odenbach2003ferrofluids}
S.~Odenbach.
\newblock Ferrofluids--magnetically controlled suspensions.
\newblock {\em Colloid Surface A}, 217(1--3):171--178, 2003.

\bibitem{odenbach2003magnetoviscous}
S.~Odenbach.
\newblock {\em Magnetoviscous effects in ferrofluids}.
\newblock Springer Berlin / Heidelberg, 2003.

\bibitem{odenbach2004recent}
S.~Odenbach.
\newblock Recent progress in magnetic fluid research.
\newblock {\em J. Phys.: Condens. Matter}, 16:R1135--R1150, 2004.

\bibitem{ilg2005structure}
P.~Ilg, M.~Kr{\"o}ger, and S.~Hess.
\newblock Structure and rheology of model-ferrofluids under shear flow.
\newblock {\em J. Magn. Magn. Mater.}, 289:325--327, 2005.

\bibitem{ilg2006structure}
P.~Ilg, E.~Coquelle, and S.~Hess.
\newblock Structure and rheology of ferrofluids: simulation results and kinetic
  models.
\newblock {\em J. Phys.: Condens. Matter}, 18(38):S2757--S2770, 2006.

\bibitem{holm2005structure}
C.~Holm and J.-J.~Weis.
\newblock The structure of ferrofluids: A status report.
\newblock {\em Curr. Opin. Colloid Interface Sci.}, 10(3):133--140, 2005.

\bibitem{menzel2014tuned}
A.~M.~Menzel.
\newblock Tuned, driven, and active soft matter.
\newblock {\em Phys. Rep.}, 554(0):1--45, 2014.

\bibitem{jarkova2003hydrodynamics}
E.~Jarkova, H.~Pleiner, H.-W.~M{\"u}ller, and H.~R. Brand.
\newblock Hydrodynamics of isotropic ferrogels.
\newblock {\em Phys. Rev. E}, 68(4):041706, 2003.

\bibitem{zrinyi1996deformation}
M.~Zr{\'\i}nyi, L.~Barsi, and A.~B{\"u}ki.
\newblock Deformation of ferrogels induced by nonuniform magnetic fields.
\newblock {\em J. Chem. Phys.}, 104(21):8750--8756, 1996.

\bibitem{deng2006development}
H.-X. Deng, X.-L. Gong, and L.-H. Wang.
\newblock Development of an adaptive tuned vibration absorber with
  magnetorheological elastomer.
\newblock {\em Smart Mater. Struct.}, 15(5):N111--N116, 2006.

\bibitem{stepanov2007effect}
G.~V. Stepanov, S.~S. Abramchuk, D.~A. Grishin, L.~V. Nikitin, E.~Y.
  Kramarenko, and A.~R. Khokhlov.
\newblock Effect of a homogeneous magnetic field on the viscoelastic behavior
  of magnetic elastomers.
\newblock {\em Polymer}, 48(2):488--495, 2007.

\bibitem{filipcsei2007magnetic}
G.~Filipcsei, I.~Csetneki, A.~Szil{\'a}gyi, and M.~Zr\'{i}nyi.
\newblock Magnetic field-responsive smart polymer composites.
\newblock {\em Adv. Polym. Sci.}, 206:137--189, 2007.

\bibitem{guan2008magnetostrictive}
X.~Guan, X.~Dong, and J.~Ou.
\newblock Magnetostrictive effect of magnetorheological elastomer.
\newblock {\em J. Magn. Magn. Mater.}, 320(3--4):158--163, 2008.

\bibitem{bose2009magnetorheological}
H.~B{\"o}se and R.~R{\"o}der.
\newblock Magnetorheological elastomers with high variability of their
  mechanical properties.
\newblock {\em J. Phys.: Conf. Ser.}, 149(1):012090, 2009.

\bibitem{gong2012full}
X.~Gong, G.~Liao, and S.~Xuan.
\newblock Full-field deformation of magnetorheological elastomer under uniform
  magnetic field.
\newblock {\em Appl. Phys. Lett.}, 100(21):211909, 2012.

\bibitem{evans2012highly}
B.~A. Evans, B.~L. Fiser, W.~J. Prins, D.~J. Rapp, A.~R. Shields, D.~R. Glass,
  and R.~Superfine.
\newblock A highly tunable silicone-based magnetic elastomer with nanoscale
  homogeneity.
\newblock {\em J. Magn. Magn. Mater.}, 324(4):501--507, 2012.

\bibitem{borin2013tuning}
D.~Y. Borin, G.~V. Stepanov, and S.~Odenbach.
\newblock Tuning the tensile modulus of magnetorheological elastomers with
  magnetically hard powder.
\newblock {\em J. Phys.: Conf. Ser.}, 412(1):012040, 2013.

\bibitem{zimmermann2006modelling}
K.~Zimmermann, V.~A. Naletova, I.~Zeidis, V.~B{\"o}hm, and E.~Kolev.
\newblock Modelling of locomotion systems using deformable magnetizable media.
\newblock {\em J. Phys.: Condens. Matter}, 18(38):S2973--S2983, 2006.

\bibitem{szabo1998shape}
D.~Szab\'{o}, G.~Szeghy, and M.~Zr\'{i}nyi.
\newblock Shape transition of magnetic field sensitive polymer gels.
\newblock {\em Macromolecules}, 31(19):6541--6548, 1998.

\bibitem{ramanujan2006mechanical}
R.~V. Ramanujan and L.~L. Lao.
\newblock The mechanical behavior of smart magnet-hydrogel composites.
\newblock {\em Smart Mater. Struct.}, 15(4):952--956, 2006.

\bibitem{sun2008study}
T.~L. Sun, X.~L. Gong, W.~Q. Jiang, J.~F. Li, Z.~B. Xu, and W.H. Li.
\newblock Study on the damping properties of magnetorheological elastomers
  based on cis-polybutadiene rubber.
\newblock {\em Polym. Test.}, 27(4):520--526, 2008.

\bibitem{babincova2001superparamagnetic}
M.~Babincov{\'a}, D.~Leszczynska, P.~Sourivong, P.~{\v{C}}i{\v{c}}manec, and
  P.~Babinec.
\newblock Superparamagnetic gel as a novel material for electromagnetically
  induced hyperthermia.
\newblock {\em J. Magn. Magn. Mater.}, 225(1):109--112, 2001.

\bibitem{lao2004magnetic}
L.~L. Lao and R.~V. Ramanujan.
\newblock Magnetic and hydrogel composite materials for hyperthermia
  applications.
\newblock {\em J. Mater. Sci.: Mater. Med.}, 15(10):1061--1064, 2004.

\bibitem{landau1975elasticity}
L.~D. Landau and E.~M. Lifshitz.
\newblock {\em Elasticity theory}.
\newblock Pergamon Press, 1975.

\bibitem{bohlius2004macroscopic}
S.~Bohlius, H.~R. Brand, and H.~Pleiner.
\newblock Macroscopic dynamics of uniaxial magnetic gels.
\newblock {\em Phys. Rev. E}, 70(6):061411, 2004.

\bibitem{annunziata2013hardening}
M.~A. Annunziata, A.~M. Menzel, and H. L{\"o}wen.
\newblock Hardening transition in a one-dimensional model for ferrogels.
\newblock {\em J. Chem. Phys.}, 138(20):204906, 2013.

\bibitem{messing2011cobalt}
R.~Messing, N.~Frickel, L.~Belkoura, R.~Strey, H.~Rahn, S.~Odenbach, and A.~M.
  Schmidt.
\newblock Cobalt ferrite nanoparticles as multifunctional cross-linkers in paam
  ferrohydrogels.
\newblock {\em Macromolecules}, 44(8):2990--2999, 2011.

\bibitem{zubarev2012theory}
A.~Y. Zubarev.
\newblock On the theory of the magnetic deformation of ferrogels.
\newblock {\em Soft Matter}, 8(11):3174--3179, 2012.

\bibitem{brand2014macroscopic}
H.~R. Brand and H.~Pleiner.
\newblock Macroscopic behavior of ferronematic gels and elastomers.
\newblock {\em Eur. Phys. J. E}, 37(12):122, 2014.

\bibitem{ivaneyko2012effects}
D.~Ivaneyko, V.~Toshchevikov, M.~Saphiannikova, and G.~Heinrich.
\newblock Effects of particle distribution on mechanical properties of
  magneto-sensitive elastomers in a homogeneous magnetic field.
\newblock {\em Condens. Matter Phys.}, 15(3):33601, 2012.

\bibitem{wood2011modeling}
D.~S. Wood and P.~J. Camp.
\newblock Modeling the properties of ferrogels in uniform magnetic fields.
\newblock {\em Phys. Rev. E}, 83(1):011402, 2011.

\bibitem{pessot2014structural}
G.~Pessot, P.~Cremer, D.~Y. Borin, S.~Odenbach, H.~L{\"o}wen, and A.~M Menzel.
\newblock Structural control of elastic moduli in ferrogels and the importance
  of non-affine deformations.
\newblock {\em J. Chem. Phys.}, 141(12):124904, 2014.

\bibitem{raikher2008shape}
Y.~L. Raikher, O.~V. Stolbov, and G.~V. Stepanov.
\newblock Shape instability of a magnetic elastomer membrane.
\newblock {\em J. Phys. D}, 41:152002, 2008.

\bibitem{stolbov2011modelling}
O.~V. Stolbov, Y.~L. Raikher, and M.~Balasoiu.
\newblock Modelling of magnetodipolar striction in soft magnetic elastomers.
\newblock {\em Soft Matter}, 7(18):8484--8487, 2011.

\bibitem{han2013field}
Y.~Han, W.~Hong, and L.~E. Faidley.
\newblock Field-stiffening effect of magneto-rheological elastomers.
\newblock {\em Int. J. Solids Struct.}, 50(14--15):2281--2288, 2013.

\bibitem{spieler2013xfem}
C.~Spieler, M.~K{\"a}stner, J.~Goldmann, J.~Brummund, and V.~Ulbricht.
\newblock Xfem modeling and homogenization of magnetoactive composites.
\newblock {\em Acta Mech.}, 224(11):2453--2469, 2013.

\bibitem{dudek2007molecular}
M.~R Dudek, B.~Grabiec, and K.~W. Wojciechowski.
\newblock Molecular dynamics simulations of auxetic ferrogel.
\newblock {\em Rev. Adv. Mater. Sci}, 14:167--173, 2007.

\bibitem{sanchez2013effects}
P.~A. S\'{a}nchez, J.~J. Cerd\`{a}, T.~Sintes, and C.~Holm.
\newblock Effects of the dipolar interaction on the equilibrium morphologies of
  a single supramolecular magnetic filament in bulk.
\newblock {\em J. Chem. Phys.}, 139(4):044904, 2013.

\bibitem{tarama2014tunable}
M.~Tarama, P.~Cremer, D.~Y. Borin, S.~Odenbach, H.~L{\"o}wen, and A.~M. Menzel.
\newblock Tunable dynamic response of magnetic gels: Impact of structural
  properties and magnetic fields.
\newblock {\em Physical Review E}, 90(4):042311, 2014.

\bibitem{cerda2013phase}
J.~J. {Cerd{\`a}}, P.~A. {S{\'a}nchez}, C.~{Holm}, and T.~{Sintes}.
\newblock Phase diagram for a single flexible stockmayer polymer at zero field.
\newblock {\em Soft Matter}, 9:7185, 2013.

\bibitem{weeber2012deformation}
R.~Weeber, S.~Kantorovich, and C.~Holm.
\newblock Deformation mechanisms in 2d magnetic gels studied by computer
  simulations.
\newblock {\em Soft Matter}, 8:9923--9932, 2012.

\bibitem{weeber2015ferrogels}
R.~Weeber, S.~Kantorovich, and C.~Holm.
\newblock Ferrogels cross-linked by magnetic nanoparticles-deformation
  mechanisms in two and three dimensions studied by means of computer
  simulations.
\newblock {\em J. Magn. Magn. Mater.}, 383(0):262 -- 266, 2015.

\bibitem{harmandaris2007comparison}
V.~A. Harmandaris, D.~Reith, N.~F.~A.~van~der~Vegt, and K.~Kremer.
\newblock Comparison between coarse-graining models for polymer systems: Two
  mapping schemes for polystyrene.
\newblock {\em Macromol. Chem. Phys.}, 208(19-20):2109--2120, 2007.

\bibitem{mulder2008equilibration}
T.~Mulder, V.~A. Harmandaris, A.~V. Lyulin, N.~F.~A.~van~der~Vegt, B.~Vorselaars, and M.~A.~J.~Michels.
\newblock Equilibration and deformation of amorphous polystyrene: Scale-jumping
  simulational approach.
\newblock {\em Macromol. Theory Simul.}, 17(6):290--300, 2008.

\bibitem{menzel2014bridging}
A.~M. Menzel.
\newblock Bridging from particle to macroscopic scales in uniaxial magnetic
  gels.
\newblock {\em J. Chem. Phys.}, 141(19):194907, 2015.

\bibitem{arnold13a}
A.~Arnold, O.~Lenz, S.~Kesselheim, R.~Weeber, F.~Fahrenberger, D.~R{\"o}hm,
  P.~Ko\v{s}ovan, and C.~Holm.
\newblock Espresso 3.1: Molecular dynamics software for coarse-grained models.
\newblock In Michael Griebel and Marc~Alexander Schweitzer, editors, {\em
  Meshfree Methods for Partial Differential Equations VI}, volume~89 of {\em
  Lecture Notes in Computational Science and Engineering}, pages 1--23.
  Springer Berlin Heidelberg, 2013.

\bibitem{hansen2002effective}
J.~P. Hansen and H.~L{\"o}wen.
\newblock Effective interactions for large-scale simulations of complex fluids.
\newblock In Peter Nielaba, Michel Mareschal, and Giovanni Ciccotti, editors,
  {\em Bridging Time Scales: Molecular Simulations for the Next Decade}, volume
  605 of {\em Lecture Notes in Physics}, pages 167--196. Springer Berlin
  Heidelberg, 2002.

\bibitem{limbach06a}
H.~J. Limbach, A.~Arnold, B.~A. Mann, and C.~Holm.
\newblock {ESPResSo} -- an extensible simulation package for research on soft
  matter systems.
\newblock {\em Comp. Phys. Comm.}, 174(9):704--727, May 2006.

\bibitem{wang2002molecular}
Z.~Wang, C.~Holm, and H.~W.~M{\"u}ller.
\newblock Molecular dynamics study on the equilibrium magnetization properties
  and structure of ferrofluids.
\newblock {\em Phys. Rev. E}, 66:021405, 2002.

\bibitem{frenkel02b}
D.~Frenkel and B.~Smit.
\newblock {\em Understanding Molecular Simulation}.
\newblock Academic Press, San Diego, second edition, 2002.

\bibitem{frickel2011magneto}
N.~Frickel, R.~Messing, and A.~M. Schmidt.
\newblock Magneto-mechanical coupling in
  \uppercase{C}o\uppercase{F}e$_2$\uppercase{O}$_4$-linked \uppercase{PAA}m
  ferrohydrogels.
\newblock {\em J. Mater. Chem.}, 21(23):8466--8474, 2011.

\bibitem{warner1972kinetic}
H.~R. Warner.
\newblock Kinetic theory and rheology of dilute suspensions of finitely
  extendible dumbbells.
\newblock {\em Ind. Eng. Chem. Fundam.}, 11(3):379--387, 1972.

\bibitem{suppl}
\am{Corresponding data curves from the additional microscopic simulations and fitted mesoscopic model curves are summarized in the supplemental material.}

\bibitem{nelder1965simplex}
J.~A. Nelder and R.~Mead.
\newblock A simplex method for function minimization.
\newblock {\em Comput. J.}, 7(4):308--313, 1965.

\bibitem{jones2001scipy}
E.~Jones, T.~Oliphant, and P.~Peterson.
\newblock {SciPy}: Open source scientific tools for {Python}, 2001--.

\bibitem{gunther2012xray}
D.~G{\"u}nther, D.~Y. Borin, S.~G{\"u}nther, and S.~Odenbach.
\newblock X-ray micro-tomographic characterization of field-structured
  magnetorheological elastomers.
\newblock {\em Smart Mater. Struct.}, 21(1):015005, 2012.

\bibitem{collin2003frozen}
D.~Collin, G.~K. Auernhammer, O.~Gavat, P.~Martinoty, and H.~R. Brand.
\newblock Frozen-in magnetic order in uniaxial magnetic gels: preparation and
  physical properties.
\newblock {\em Macromol. Rapid Commun.}, 24(12):737--741, 2003.

\bibitem{varga2003smart}
Z.~Varga, J.~Feh\'{e}r, G.~Filipcsei, and M.~Zr{\'\i}nyi.
\newblock Smart nanocomposite polymer gels.
\newblock {\em Macromol. Symp.}, 200(1):93--100, 2003.

\bibitem{borbath2012xmuct}
T.~Borb{\'a}th, S.~G{\"u}nther, D.~Y. Borin, T. Gundermann, and S.~Odenbach.
\newblock X$\mu$\uppercase{CT} analysis of magnetic field-induced phase
  transitions in magnetorheological elastomers.
\newblock {\em Smart Mater. Struct.}, 21(10):105018, 2012.

\end{thebibliography}

\end{document}